\documentclass[a4paper,fleqn]{arxiv}

\usepackage{graphicx}
\usepackage{epstopdf, epsfig}
\usepackage{booktabs}
\usepackage{multirow}
\usepackage{array}
\usepackage{calc}
\usepackage{microtype}
\usepackage{lscape}
\usepackage{esint,upgreek}
\usepackage{amsmath,amsfonts}
\usepackage{bm}
\usepackage{afterpage}

\usepackage{framed} 
\usepackage{multicol} 
\usepackage{nomencl} 
\makenomenclature
\renewcommand*\nompreamble{\begin{multicols}{2}}
\renewcommand*\nompostamble{\end{multicols}}

\usepackage{algorithm2e}
\RestyleAlgo{ruled}

\usepackage[normalem]{ulem}
\usepackage{color}
\usepackage{mylines}
\usepackage{myplots}
\newcommand{\sy}[2]{\mbox{(\kern-.25em\SymbolRGB[solid]{#1}{.8pt}{#2}{5pt}\kern-.25em)}}
\newcommand{\lsy}[3]{\mbox{(\kern-.1em\lineSymbolRGB{#1}{#2}{2pt}{#3}{4pt}\kern-.45em)}}
\newcommand{\lcap}[2]{~\,{\kern-1em\protect\mylcap{#1}{#2}}}

\definecolor{blue}{rgb}{0,0,1}
\definecolor{red}{rgb}{1,0,0}
\definecolor{black}{rgb}{0,0,0}
\definecolor{grey}{rgb}{0.775,0.775,0.775}
\definecolor{white}{rgb}{1,1,1}

\definecolor{BlueTBL1}{rgb}{0.41961,0.682353,0.839216}
\definecolor{BlueTBL2}{rgb}{0.12941,0.443137,0.709804}
\definecolor{green_laser}{RGB}{0,208,0}
\definecolor{green_lgac}{rgb}{0.1068,0.5190,0.2488}
\definecolor{red_lgac}{rgb}{0.7630,0.0863,0.1068}
\definecolor{blue_lgac}{rgb}{0.1076,0.4153,0.6880}
\definecolor{red_pod_1}{rgb}{0.9467,0.2682,0.1961}
\definecolor{red_pod_2}{rgb}{0.7365,0.0800,0.1012}
\definecolor{red_pod_3}{rgb}{0.4039,0,0.0510}

\usepackage[numbers]{natbib}

\def\tsc#1{\csdef{#1}{\textsc{\lowercase{#1}}\xspace}}
\tsc{WGM}
\tsc{QE}

\begin{document}
\let\WriteBookmarks\relax
\def\floatpagepagefraction{1}
\def\textpagefraction{.001}

\shorttitle{Genetically-inspired convective heat transfer enhancement}

\shortauthors{R. Castellanos et al.}  

\title [mode = title]{Genetically-inspired convective heat transfer enhancement in a turbulent boundary layer}

\author[1,2]{Rodrigo Castellanos}[
    orcid=0000-0002-7789-5725]
\cormark[1] 
\ead{rcastell@ing.uc3m.es}
\credit{Conceptualization, Methodology, Software, Validation, Formal analysis, Investigation, Data Curation, Writing - Original Draft, Writing - Review \& Editing, Visualization, Project administration}
\affiliation[1]{organization={Aerospace Engineering Research Group, Universidad Carlos III de Madrid},
            city={Legan\'es},
            postcode={28911}, 
            state={Madrid},
            country={Spain}}
\affiliation[2]{organization={Theoretical and Computational Aerodynamics Branch, Flight Physics Department, Spanish National Institute for Aerospace Technology (INTA)},
            city={Torrej\'on de Ardoz},
            postcode={28850}, 
            state={Madrid},
            country={Spain}}
\author[1]{Andrea Ianiro}[
    orcid=0000-0001-7342-4814]
\ead{aianiro@ing.uc3m.es}
\credit{Conceptualization, Methodology, Validation, Resources, Writing - Review \& Editing, Supervision, Project administration, Funding acquisition}
\author[1]{Stefano Discetti}[
    orcid=0000-0001-9025-1505]
\ead{sdiscett@ing.uc3m.es}
\credit{Conceptualization, Methodology, Software, Validation, Resources, Writing - Review \& Editing, Supervision, Project administration, Funding acquisition}
\cortext[1]{Corresponding author}

\maketitle

\noindent \textsc{Abstract}\\
\\
The convective heat transfer in a turbulent boundary layer (TBL) on a flat plate is enhanced using an artificial intelligence approach based on linear genetic algorithms control (LGAC). The actuator is a set of six slot jets in crossflow aligned with the freestream. An open-loop optimal periodic forcing is defined by the carrier frequency, the duty cycle and the phase difference between actuators as control parameters. The control laws are optimised with respect to the unperturbed TBL and to the actuation with a steady jet. The cost function includes the wall convective heat transfer rate and the cost of the actuation. The performance of the controller is assessed by infrared thermography and characterised also with particle image velocimetry measurements. The optimal controller yields a slightly asymmetric flow field. The LGAC algorithm converges to the same frequency and duty cycle for all the actuators. It is noted that such frequency is strikingly equal to the inverse of the characteristic travel time of large-scale turbulent structures advected within the near-wall region. The phase difference between multiple jet actuation has shown to be very relevant and the main driver of flow asymmetry. The results pinpoint the potential of machine learning control in unravelling unexplored controllers within the actuation space. Our study furthermore demonstrates the viability of employing sophisticated measurement techniques together with advanced algorithms in an experimental investigation.

\vspace{1cm}
\noindent \textsc{Keywords}\\
\\
Machine learning \\ Genetic algorithms \\ Flow control \\ Pulsed crossflow jets \\ Turbulent Boundary layers \\ Convective heat transfer enhancement

\begin{table*}[hbt]   
    \begin{framed}
    \nomenclature[01]{$\alpha,\beta$}{Diagnostic-plot parameters}
    \nomenclature[02]{$\beta_v$}{Volumetric thermal expansion coefficient}
    \nomenclature[03]{$\gamma$}{Penalisation coefficient}
    \nomenclature[04]{$\delta$}{Boundary layer thickness}
    \nomenclature[05]{$\delta^*$}{Boundary layer displacement thickness} 
    \nomenclature[06]{$\varepsilon$}{Emissivity}
    \nomenclature[07]{$\theta$}{Boundary layer momentum thickness} 
    \nomenclature[08]{$\kappa(~)$}{Proportional scaling} 
    \nomenclature[09]{$\nu$}{Kinematic viscosity}
    \nomenclature[10]{$\Xi$}{Actuation parameters}
    \nomenclature[11]{$\xi$}{Uncertainty of infrared thermography data}
    \nomenclature[12]{$\rho$}{Density}
    \nomenclature[13]{$\sigma$}{Boltzmann constant}
    \nomenclature[14]{$\tau$}{Pulse width of the actuation}
    \nomenclature[15]{$\tau_{open}$}{Opening lapse time of the valve}
    \nomenclature[16]{$\tau_{close}$}{Closing lapse time of the valve}
    \nomenclature[17]{$\phi$}{Phase of the actuation}
    
    \nomenclature[18]{$A_\mathrm{foil}$}{Area of the heated-thin-foil sensor}
    \nomenclature[19]{$a_g$}{Acceleration of gravity}
    \nomenclature[20]{$\mathrm{Bi}$}{Biot number}
    \nomenclature[21]{$\mathcal{B}$}{Space of actuation commands}
    \nomenclature[22]{$b$}{Actuation command}
    \nomenclature[23]{$c$}{Convection velocity ($\approx10\cdot u_\tau$)}
    \nomenclature[24]{$DC$}{Duty cycle of the actuation}
    \nomenclature[26]{$E$}{Convective heat transfer enhancement}
    \nomenclature[27]{$e_r$}{Relative error for repetition $r$}
    \nomenclature[28]{$f$}{Carrier frequency of the actuation}
    \nomenclature[29]{$f\#$}{Focal ratio}
    \nomenclature[30]{$\mathrm{Gr}$}{Grashoff number}
    \nomenclature[31]{$g$}{Generation index}
    \nomenclature[32]{$H(~)$}{Heaviside function}
    \nomenclature[32]{$H_{12}$}{Boundary layer shape factor}
    \nomenclature[33]{$h$}{Convective heat transfer coefficient}
    \nomenclature[34]{$h_b$}{$h$ on the back side of the sensor}
    \nomenclature[35]{$I$}{Electric current}
    \nomenclature[36]{$i$}{Individual index}
    \nomenclature[37]{$J$}{Cost function}
    \nomenclature[38]{$J_a$}{Cost associated to heat transfer enhancement}
    \nomenclature[39]{$J_b$}{Cost associated to mass-flow injection}
    \nomenclature[40]{$K$}{Controller}
    \nomenclature[41]{$\mathbb{K}$}{Space of control laws}
    \nomenclature[42]{$k_i$}{Thermal conductivity of element $i$}
    \nomenclature[43]{$L$}{Reference/characteristic length}
    \nomenclature[44]{$\dot{m}$}{Mass flow rate}
    \nomenclature[45]{$N_b$}{Number of actuation commands}
    \nomenclature[46]{$N_g$}{Number of generations}
    \nomenclature[47]{$N_p$}{Number of actuation parameters}
    \nomenclature[48]{$\mathrm{Nu}$}{Nusselt number}
    \nomenclature[49]{$P$}{Pumping power}
    \nomenclature[50]{$\mathcal{P}$}{Space of the controller's parameters}
    \nomenclature[52]{$q''_{b}$}{Heat flux through the back side of the sensor}
    \nomenclature[53]{$q''_{j}$}{Heat flux by Joule effect}
    \nomenclature[54]{$q''_{k}$}{Heat flux by conduction}
    \nomenclature[55]{$q''_{r}$}{Heat flux by radiation}
    \nomenclature[56]{$R$}{Velocity ratio}
    \nomenclature[57]{$\mathrm{Re}$}{Reynolds number}
    \nomenclature[58]{$\mathrm{Re}_\tau$}{Friction-based Reynolds number}
    \nomenclature[59]{$\mathrm{Re}_\theta$}{Reynolds number based on momentum thickness}
    \nomenclature[60]{$r$}{repetition index}
    \nomenclature[61]{$\mathrm{St}$}{Strouhal number}
    \nomenclature[62]{$T$}{Actuation period}
    \nomenclature[63]{$T_\mathrm{w}$}{Wall temperature}
    \nomenclature[64]{$T_\mathrm{aw}$}{Adiabatic wall temperature}
    \nomenclature[65]{$T_\infty$}{Freestream temperature}
    \nomenclature[66]{$t_1$}{Convergence time for $T_\mathrm{aw}$}
    \nomenclature[67]{$t_2$}{Data acquisition time for $T_\mathrm{aw}$}
    \nomenclature[68]{$t_3$}{Convergence time for $T_\mathrm{w}$}
    \nomenclature[69]{$t_4$}{Data acquisition time for $T_\mathrm{w}$}
    \nomenclature[70]{$t_i$}{Thickness of element $i$}
    \nomenclature[71]{$U,V,W$}{Streamwise, wall-normal,and velocity}
    \nomenclature[72]{$U_\infty$}{Freestream velocity}
    \nomenclature[73]{$U_\mathrm{jet}$}{Jet velocity}
    \nomenclature[74]{$u,v,w$}{Streamwise, wall-normal,and velocity fluctuations}
    \nomenclature[75]{$u_\tau$}{Friction velocity}
    \nomenclature[76]{$V_{DC}$}{Electric voltage}
    \nomenclature[77]{$x,y,z$}{Streamwise, wall-normal,and spanwise directions}
    \nomenclature[78]{$x_\mathrm{jet}$}{Streamwise position of the jets}
    \nomenclature[79]{$x_s$}{Upstream edge of the heat-flux sensor}
    \nomenclature[80]{$\hat{x}$}{$=x-x_s$}
    \nomenclature[81]{$\widetilde{\bullet}$}{Normalise with the unperturbed condition operator}
    \nomenclature[82]{$\overline{\hspace{0.5mm} \bullet \hspace{0.5mm}}$}{Time-average operator}
    \nomenclature[83]{$\langle{\bullet}\rangle_{x,z}$}{Streamwise-, spanwise-average operator}
    \nomenclature[84]{${+}$}{Superscript for inner units}
    \nomenclature[85]{${*}$}{Superscript for optimised controller}
    \nomenclature[86]{${\star}$}{Superscript for best individual}
    \nomenclature[87]{${\mathrm{sj}}$}{Subscript for steady-jet condition}
    \nomenclature[88]{${0}$}{Subscript for unperturbed condition}
    
    \printnomenclature
    \end{framed}
\end{table*}

\newpage
\section{Introduction}\label{s:intro}
The control of turbulent flows based on artificial intelligence has gained interest due to its versatility and superlative capabilities to deal with the complexity imposed by the non-linearity, time-dependence, and high dimensionality inherent to the Navier-Stokes equations \citep{BruntonNoackKoumoutsakos2020}. In this work, we enhance the convective heat transfer in a turbulent boundary layer (TBL) using a model-free self-learning control method. 

Convective heat transfer enhancement in turbulent boundary layer flows has been achieved with passive solutions, such as obstacles \citep{mallor2018cubes} or riblets \citep{mallor2019modal} and vortex generators \citep{Ke2019VG}, and active devices like continuous \citep{puzu2019jet}, synthetic \citep{Giachetti2018synjet}, and pulsed \citep{Castellanos2022slotjet} jets in crossflow, plasma actuators \citep{roy2008plasma}, or electromagnetic forcing \citep{Kenjere2008EMheat}, among others. In this study, the actuator consists of a spanwise array of pulsed slot jets in cross flow (JICF).  The flow physics and the induced structures by JICF are well described in the literature \citep[e.g.][]{cortelezzi2001formation, Getsinger2014}. Jet pulsation can be exploited to optimise a certain flow feature \citep{eroglu2001structure, MCLOSKEY2002, shapiro2006optimization}. For instance, \citet{johari1999penetration} studied the mixing properties of fully-modulated JICF, finding that the frequency and duty cycle are the main control parameters influencing the mixing properties. 

Most of the studies employing JICF for flow control exploit the formation of a counter-rotating vortex pair for mixing enhancement. Conventionally, the generation of embedded streamwise vortices in TBLs is also considered one of the main heat transfer enhancement mechanisms \citep{jacobi1995}. The numerical study by \citet{Zhang1993heatJICF} concludes that the heat transfer increases due to the entrained fluid towards the wall provided by the induced counter-rotating vortex pair. Likewise, the recent effort by \citet{puzu2019jet} investigates the influence of the jet-induced counter-rotating vortices in the enhancement of heat transfer. The interested reader is referred to the extensive review article by \citet{karagozian2010transverse} covering the jet in crossflow and the control based on it. 

\afterpage{\clearpage} 

Effective turbulence control is a challenging task that mainly depends on the phenomena to be controlled and the available information from the system. The flow control problem is commonly posed as an optimisation problem. Active control based on periodic forcing, for instance, may be parameterised by the most relevant control parameters such as the pulsation frequency, the duty cycle or the phase shift between actuators. Some examples are the skin-friction reduction in a TBL by periodic blowing and suction \citep{cheng2021skin}, the periodic pulsation of a jet in cross-flow to enhance heat transfer in a TBL \citep{Castellanos2022slotjet} or mixing in a jet stream \citep{Fan2017jetcontrol}. The performance of the periodic control scales with the size of the feasible solution space; however, the optimisation process becomes more time-consuming. 

The irruption of artificial intelligence algorithms opens new knowledge gaps in which the usage of more sophisticated optimisation methodologies can lead to outstanding performance enhancements \citep{Noack2019control}. There is a vast collection of mathematical tools for an effective design of the control law, distinguishing between model-based and model-free control depending on whether a model to describe the dynamics is available.

Model-based control stands out when dealing with simplified flow-control problems based on first- or second-order dynamics \citep{Rowley2006control}. Model-based control strategies enable a profound understanding of the dynamics of the actuated system based on a pre-established model. This approach has hitherto struggled in controlling turbulent flows due to the chaotic dynamics of turbulence with countless frequency-crosstalk mechanisms \citep{BruntonNoack2015review}. A few exceptions succeeded in controlling elementary turbulent flows, mainly exploiting cases in which the linear dynamics can be numerically determined by discretising the Navier-Stokes (N–S) equations  \citep{Kim2007linearcontrol,Sipp2010linearcontrol}. The creation of transfer functions based on the parabolised stability equations also allows a real-time estimation of the wavepacket evolution \citep{sasaki2017wavepackets}. Some other examples related to flow control of first- and second-order dynamics include near-wall opposition control \citep{Choi1994,Fukagata2003oppcontrol},  two-frequency crosstalk \citep{Glezer2005control,luchtenburg2009galerkin}, oscillations stabilisation via phasor control \citep{pastoor2008control}, and quasi-steady response to quasi-steady actuation \citep{Pfeiffer2018robustcontrol}.

The intrinsic challenges posed by turbulence control motivate model-free approaches. Nevertheless, the non-convexity of the feasible solution space for the actuation design represents a challenge. In experimental applications, most of the literature studies simplify the optimisation problem to the identification of a few actuation parameters with adaptive variations either based on temporal evolution or sensor signal. Common model-free strategies for turbulence control are the evolutionary \citep{Koumoutsakos2001evolcontrol} and genetic algorithms \citep{Benard2016ga}. Other examples include extremum and slope-seeking control methods \citep{Krstic2000extremumseek, Becker2007extseek, Gelbert2012extseek} and physics-based methods \citep{pastoor2008control,Zhang2004control}. 

Model-free approaches have been increasingly applied in the last decade, thanks to the popularisation and advances in machine learning techniques. An early precursor is the work by \citet{Lee1997NN} pioneering neural network control for skin-friction drag reduction. Reinforcement Learning (RL) has recently become one of the most prominent model-free control techniques from the machine learning literature \citep{rabault2019JFM,Beintema2020,li2021ReLe,paris2021}. On the other hand, evolutionary algorithms \citep{Dracopoulos1997geneticalg}, rediscovered as machine learning control (MLC), turn up as a promising alternative in flow control applications \citep{duriez2017book,li2017GP}. A comparative assessment of the two methods has been recently published by \citet{Castellanos2022LGPCvsRL}, testing their performance when only a reduced number of (noisy) sensors is available. 

In the current work, we maximise the convective heat transfer enhancement in a turbulent boundary layer. This is a complicated flow-control problem that comprises a wide range of spatial and temporal scales with mutual nonlinear interaction. Model-based control presents strong limitations in presence of highly-chaotic flows. The authors are unaware of any promising model-based control study applied to turbulent flows targeted to heat-transfer enhancement. Therefore, we employ MLC based on linear genetic algorithms control (LGAC) for the control logic. Evolutionary algorithms are rather unexplored in the field of heat transfer. The extensive review by \citet{Gosselin2009GareviewHT} describes most of the applications in this field. Most recently, the work by \citet{Tian2020GAheatexchanger} used a genetic algorithm (GA) to increase the efficiency of a spiral double-pipe heat exchanger, and \citet{Moon2021GA-HE} optimised the design of an ultra-compact heat exchanger with a GA. Conversely, MLC has been used in a diverse portfolio of experiments as summarised in the review by \citet{Noack2019control}.

This study focuses on the enhancement of convective heat transfer in a TBL by employing a spanwise array of six slot jets in crossflow.  We intend to exploit linear genetic algorithms to explore a wide range of actuation parameters. This approach has been used successfully in previous works targeted jet mixing enhancement \cite{zhou2020artificial,perumal2022hybrid} and is extended and tuned here for heat transfer control. The present manuscript is structured as follows. Section \ref{s:Methodology} defines the experimental apparatus, the measurement techniques and the actuation system, followed by the characterisation of the unperturbed flow conditions in Section~\ref{s:Characterisation}. The optimisation problem is formulated in Section~\ref{s:MLC} together with a description of MLC. Section~\ref{s:Results} collects the results focusing on the evolution of the training process while the discussion of the optimised control law and its effect on the flow gathers in Section~\ref{s:discussion}. Ultimately, the conclusions of the study are drawn in Section \ref{s:Conclusions}.

\section{Experimental setup and Methodology \label{s:Methodology}}

This section addresses the experimental setup and measurement techniques employed in the study. The actuation system is outlined and the setup for flow field and heat transfer measurements is presented. 

\subsection{Wind tunnel setup and flow conditions \label{ss:WTandBLcond}}%
The experimental campaign was held in the G\"ottingen-type wind tunnel at Universidad Carlos III de Madrid. The test chamber expands $1.5\mathrm{m}$ of length with a $0.4 \times 0.4\mathrm{m}^2$ cross-section. The freestream velocity is set for all the test cases at $U_\infty = 12.1\mathrm{m/s}$. The streamwise turbulence intensity is below 1\%.  

\begin{figure*}[t]
    \centering
    \includegraphics[width = 0.99\linewidth]{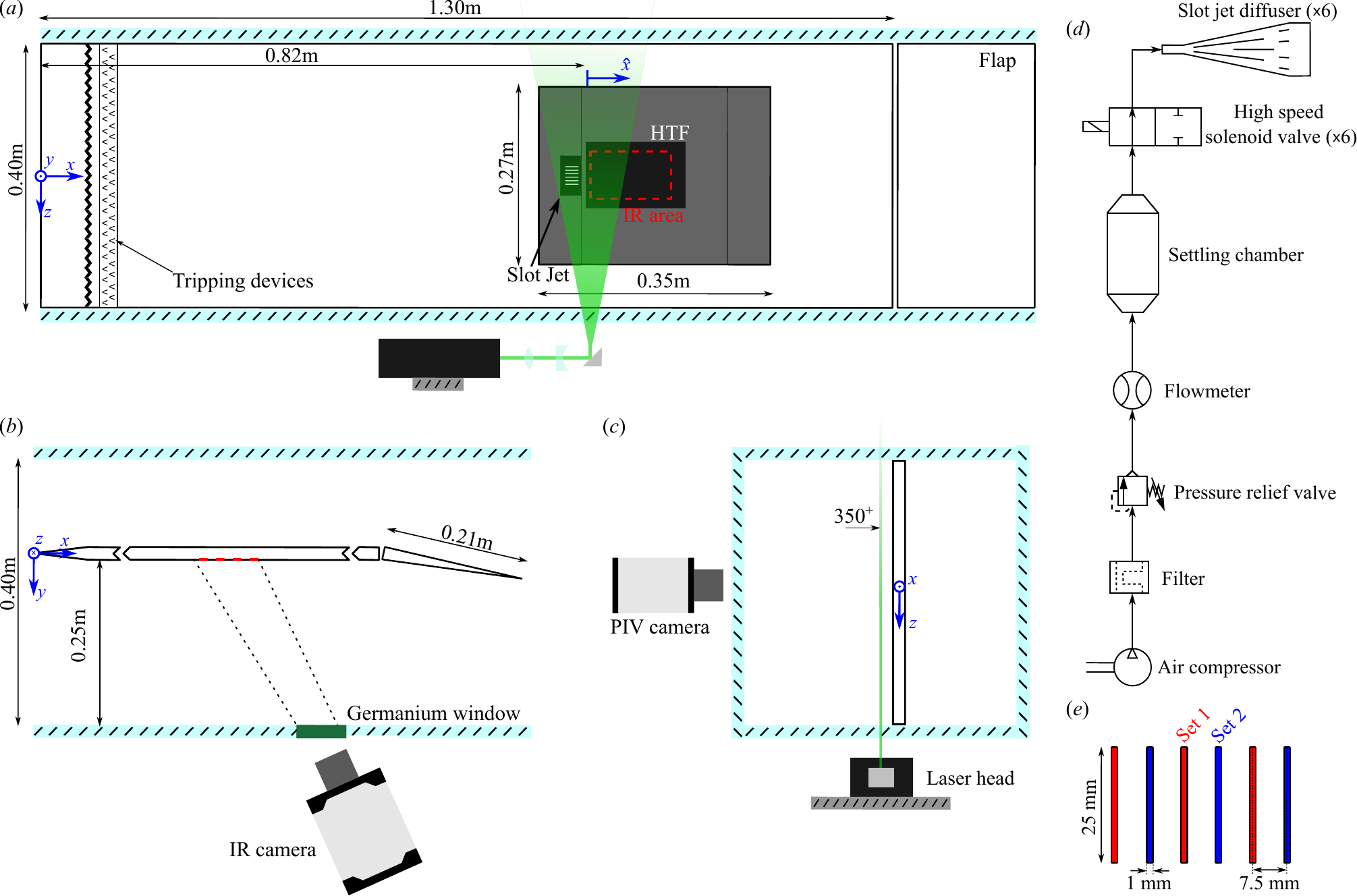}
    \caption{Schematic representation of the experimental setup in the wind tunnel: top (a), side (b) and back (c) views. The red dashed area \lcap{--}{red} indicates the IR-thermography measurement domain downstream of the actuator and the green plane \lcap{-}{green_laser} depicts the PIV illumination plane at $y^+=350$ from the wall. (d) The pneumatic system for the jet secondary flow injection. (e) Arrangement of the 6 slot jets.} \label{fig:SetUp}
\end{figure*}

The base flow consists of a turbulent boundary layer that develops on a smooth poly-methyl methacrylate flat plate of $1.5\mathrm{m}$ length and $20\mathrm{mm}$ thickness, spanning the entire width of the test section (see figure~\ref{fig:SetUp}(a-b)). The flat plate is designed to host different ad-hoc modules (grey-shaded area in figure~\ref{fig:SetUp}(a)). For this study, the plate is equipped with a 3D-printed flush-mounted module containing a spanwise array of $6$ streamwise-aligned slots and a wall heat flux sensor.

The leading edge has a wedge shape followed by a tripping device to trigger turbulence transition of the boundary layer. A $210\mathrm{mm}$ long adjustable trailing-edge flap, deflected by approximately $20^\circ$, is used to avoid flow separation at the leading edge. A combination of a $2.5\mathrm{mm}$ height zig-zag turbulators to promote an abrupt transition in combination with a $10\mathrm{mm}$ wide DYMO\textsuperscript{\textregistered} tape (with the embossed letter ‘V’ pointing in the flow direction) to diffuse the vortices shed from the upstream turbulator is used as shown in figure~\ref{fig:SetUp}(a). 

A reference system with its origin on the mid-plane of the leading edge of the flat plate is considered. For clarity, the streamwise coordinate $\hat{x}$ (see figure~\ref{fig:SetUp}(a)) will be used throughout the manuscript, defined as $\hat{x} = x - x_{s}$, being $x_{s}$ the upstream edge of the heat-flux sensor (see \S\ref{ss:IR}). Additionally, the following conventions and notations are applied. The velocity components along the streamwise $(x)$, wall-normal $(y)$ and spanwise $(z)$ directions are represented by $U$, $V$ and $W$, respectively. Over-lined symbols for velocities refer to ensemble-averaged quantities (e.g. $\overline{U}$) whereas lower-case symbols refer to the fluctuating components (e.g. $u$) according to the classical Reynolds decomposition. Wall-scaled variables are defined in terms of the local friction velocity, $u_\tau$, and kinematic viscosity, $\nu$, and are denoted by a superscript ‘+’. Other quantities used as references throughout the paper are the free-stream velocity, $U_\infty$, and the boundary-layer thickness $\delta$. 

\subsection{Actuation system \label{ss:actuator}}
%
The actuation system consists of an array of $6$ streamwise-aligned slot jets. This configuration has been adopted in recent works (see e.g. \citet{Yu2021GA_drag_slot_jets}, or the array of 31 slits used by \citet{cheng2021skin}, with drag reduction as the main target in both cases) and it is analogous to the solutions based on streamwise vortices generation with plasma actuators \cite{castellanos2022plasma,cheng2021flat}. The main hypothesis behind this choice is that spanwise travelling waves have shown a great capacity to control skin friction \citep{bai2014active}, thus we expect them to be able to tamper with convective heat transfer following Reynolds' analogy.
The actuators are placed sufficiently far downstream of the leading edge to ensure no reminiscence of the tripping devices ($x = 0.81\mathrm{m}$). The slot jets feature a rectangular cross-section of $25 \mathrm{mm}$ in length and $1 \mathrm{mm}$ in width. The slots are aligned in the streamwise direction with a centre-to-centre spacing between adjacent slots of $7.5 \mathrm{mm}$ (see figure~\ref{fig:SetUp}e). The spacing was determined based on the boundary layer thickness and the manufacturing constraints. The actual spacing corresponds to approximately one-third of the boundary layer thickness, which is a trade-off between large spacing to allow full development of the jet-induced flow downstream, and short spacing to ensure the interaction between adjacent jets. The interested reader is referred to \citet{Zhang2022slot} for an extensive analysis of the relevance of the geometric parameters of the blowing jet array.  As a figure of merit, this specific spacing coincides with the typical characteristic size of obstacles employed for convective heat transfer enhancement in TBLs \citep{mallor2018cubes,nakamura2001cube}. A diffuser is designed to ensure a smooth transition from the circular pneumatic line to the rectangular nozzle exit section. The diffuser length of $70 \mathrm{mm}$ is divided in an inlet for a rounded connection of inner diameter of $4 \mathrm{mm}$, and an expansion region of $50 \mathrm{mm}$ length with $15^\circ$ opening angle. The enlargement section is equipped with vanes to improve the flow uniformity across the slot width and reduce the pressure losses within the diffuser. The design has been carried out with the support of RANS numerical simulations.  

The pneumatic system is shared among the actuators, as sketched in figure~\ref{fig:SetUp}(d). Compressed air is filtered and regulated to a constant absolute pressure of $P_{jet} = 4.0 (\pm0.005)\mathrm{bar}$ by means of a pressure-relief valve. An Alicat Scientific\texttrademark~\textit{M-500SLPM} flow regulator controls the pressure-relief valve and monitors the flow properties such as absolute pressure, mass or volumetric flow rate, and temperature. Each jet is independently pulsed by a high-speed solenoid valve SMC\textsuperscript{\textregistered}~\textit{SX-11DJ}, with a frequency up to $350~\mathrm{Hz}$. Each solenoid valve is triggered by a $24V$ periodic square signal with a characteristic carrier frequency $f$ and duty cycle $DC$. In this study, $DC$ is defined as the ratio between the pulse width $\tau$ and the period $T$, i.e. $DC=\tau/T$. The square-wave signal driving the actuation is transferred to the valve via a ST\texttrademark~\textit{STM32 Nucleo} board with an independent timer for each valve. 

The instantaneous jet exit velocity at the centre of each slot is confirmed to be uniform longitudinally and equal for all actuators within a $<1\%$ uncertainty. The frequency response of each jet is also checked to coincide with the desired actuation frequency for the whole actuation range. It is worth noting that, according to its specifications, the SMC\textsuperscript{\textregistered}~\textit{SX-11DJ} solenoid valve has a small time delay of $0.5~\mathrm{ms}$ between the opening and closing lapse time ($\tau_{open} = 0.45\mathrm{ms}$ and $\tau_{close} = 0.40\mathrm{ms}$, respectively). This implies a progressive reduction of effective injection time as frequency increases, as reported in our previous study \citep{Castellanos2022slotjet}.

\subsection{Heat transfer measurements \label{ss:IR}}
The convective heat transfer distribution is measured using a heated-thin-foil (HTF) heat-flux sensor \citep{astarita2012infrared}. The chosen heat-flux sensor is a $20\mathrm{\mu m}$ thick stainless steel foil covering an area of $A_{\mathrm{foil}}=150\times 100\mathrm{mm}^2$ just downstream of the slot jets. The foil is flush-mounted on the flat plate model using an \textit{ad hoc}, additively manufactured frame made of polylactide (PLA). The foil lies on top of two aluminium rods covered by a Polytetrafluoroethylene (also known as Teflon) insulation. The thermally-insulated rods prevent direct contact of the foil with the PLA frame to avoid heat losses by conduction. The frame is designed to feature a $2 \mathrm{mm}$ gap between the back side of the foil and the PLA insert. 

The foil is stretched by a pair of copper clamps equipped with compressed springs to adjust the foil tension on the back side of the plate. The electrical connection between the copper and the foil is realised employing $1\mathrm{mm}$-thick indium wire embedded within a triangular engraving in the copper clamps to minimise contact resistance and local heating. Since the Biot number ($\mathrm{Bi} = h t/k$, where $h$ is the convective heat transfer coefficient, $k$ the foil thermal conductivity and $t$ its thickness) is sufficiently small ($\mathrm{Bi} \sim 10^{-5}$), the thermal gradients across the foil thickness are neglected. 

A constant heat flux by Joule effect $q''_{j}$ is provided to the HTF, supplying a stabilised current $I \approx 10.2\mathrm{A}$ and voltage $V_{DC} \approx 1.3\mathrm{V}$ ($q''_{j} = V_{DC} I / A_{\mathrm{foil}}$) through the copper clamps. Natural convection on the flow-facing side of the foil is neglected since $\mathrm{Gr/Re}^2<< 1$ ($\mathrm{Gr} = {a_g\beta_v(T_\mathrm{w}-T_\mathrm{aw}) L^3}/{\nu^2}$ is the Grashoff number, where $\beta_v$ is the volumetric thermal expansion coefficient, $a_g$ is the acceleration of gravity, $L$ is a reference length and $T_w, T_{aw}$ are the surface and adiabatic wall temperature, respectively). The convective heat transfer coefficient $h$ is expressed in dimensionless form, namely in terms of Nusselt number $\mathrm{Nu} = h \delta/k_{\mathrm{air}}$ (where $\delta$ is the boundary layer thickness and $k_{\mathrm{air}}$ is the air thermal conductivity). Thus, the coefficient $h$ is computed by posing a steady-state energy balance on the foil sensor:
\begin{equation}
	h = \frac{ q''_{j} - q''_{r} - q''_{k} - q''_{b} }{ T_\mathrm{w} - T_\mathrm{aw} } ,
	\label{eq:heatedthinfoil}
\end{equation}
where $q''_{r}$ is the radiation heat flux, $q''_{k}$ is the tangential conduction heat flux through the foil, and $q''_{b}$ is the heat flux conducted through the thin air layer and the PLA substrate on the back side of the heated-thin-foil sensor. The radiation heat flux is estimated assuming that the environment behaves as a black body at a temperature equal to that of the freestream, so that $q''_{r} = \sigma \varepsilon (T_\mathrm{w}^4 - T_\infty^4)$ (where $\sigma$ is the Boltzmann constant and $\varepsilon$ is the emissivity of the foil surface). This term is approximately $0.02 q''_{j}$ for all tested cases. The losses due to tangential condition are calculated by taking the Laplacian of the temperature on the surface of the HTF, $q''_{k} = -kt \nabla^{2} T_\mathrm{w}$.
Nevertheless, the contribution of this term is rather low, not exceeding $1\%$ of the input heat flux. Ultimately, the heat-flux losses at the back side of the foil are estimated by considering the conduction through the air gap and the PLA substrate, $q''_{b} = \left(t_{\mathrm{air}}/k_{\mathrm{air}}+t_{\mathrm{PLA}}/k_{\mathrm{PLA}}+1/h_b\right)^{-1}(T_\mathrm{w}-T_\mathrm{aw})$, being $t_m$ and $k_m$ the thickness and thermal conductivity of material $m$, and $h_b$ the convective heat transfer coefficient on the back side of the plate assumed to be of the same order of magnitude as the average turbulent heat transfer on the unperturbed TBL. The term $q''_{b}$ is approximately $0.03 q''_{j}$ for all tested cases.

Surface temperature measurements are performed with an Infratec 8820 IR camera ($640 \times 512\mathrm{pix}$ Mercury-Cadmium-Telluride detector and Noise Equivalent Temperature Difference $< 25\mathrm{mK}$), capturing images at a frequency of $10\mathrm{Hz}$ with a spatial resolution of $4.0\mathrm{pixels/mm}$. The HTF is coated with a high-emissivity paint (a similar paint was characterised by \citet{ianiro2010measurement} with emissivity $\varepsilon = 0.95$) to ensure the accuracy of the IR temperature measurements. One of the lateral walls of the test section is adjusted to feature a Germanium window, providing optical access to the IR camera. As shown in figure~\ref{fig:SetUp}(b), the camera is inclined with respect to the wall to avoid the Narcissus effect by which the camera's detector images the reflection of itself. The correspondence between camera grey levels and temperature is established with an ex-situ calibration, performed by replicating the optical path.

The adiabatic wall temperature ($T_{aw}$) is computed from a set of $200$ \textit{cold} images acquired with no heating to the foil ($q_j''= 0$); the wall temperature ($T_{w}$) derives from a set of $200$ \textit{hot} images acquired with heating to the foil activated. The uncertainties associated with the experimental measurements are calculated by a Monte Carlo simulation \citep{minkina2009infrared}, assuming statistically-uncorrelated errors and using the uncertainty values reported in table~\ref{tab:uncertainties}. The uncertainty on the Nusselt number is estimated to be lower than $\pm 5\%$. Notwithstanding, the IR thermography measurements were conducted at least twice to ensure the repeatability of results.

\begin{table}
\centering
\caption{Uncertainty contributions for Nusselt number calculation.} \label{tab:uncertainties}
    \begin{tabular*}{\tblwidth}{@{}LLL@{}}
        \toprule
        Parameter               & Uncertainty   & Typical Value \\ \midrule
        $T_\mathrm{w}$          & 0.1 K         & 310 [K]  \\
        $T_\mathrm{aw}$         & 0.1 K         & 300 [K] \\
        $T_\infty$              & 0.1 K         & 305 [K] \\
        $V_{DC}$                & 0.2\%         & 17 [V] \\
        $I$                     & 0.2\%         & 2 [A] \\
        $\varepsilon$           & 2\%           & 0.95 \\
        $A$                     & 0.1\%         & 225 [cm$^2$] \\
        $k_{{\mathrm{air}}}$    & 1\%           & 0.0265 [W/(m K)] \\
        $k_{{\mathrm{PLA}}}$    & 5\%           & 0.150 [W/(m K)] \\
        $t_{{\mathrm{air}}}$    & 1\%           & 2.0 [mm] \\
        $t_{{\mathrm{PLA}}}$    & 1\%           & 18.0 [mm] \\
        $\delta$                & 1\%           & 23 [mm] \\
        $U_\infty$              & 3\%           & 12 m/s \\
        $q''_k$                 & 10\%          & 0.5 [mW/m$^2$] \\ 
        $h_b$                   & 4\%           & 55 [W/(m$^2$ K)] \\ 
        \bottomrule
    \end{tabular*}
\end{table}

\subsection{Velocity measurements \label{ss:PIV}}
Heat transfer measurements are complemented with velocity field information. Two-component Particle Image Velocimetry (PIV) measures streamwise and spanwise velocity fields in the wall-parallel plane $y^+=350$, namely $y\approx 10\mathrm{mm}$, as shown schematically in figure \ref{fig:SetUp}(c). An Andor sCMOS camera, equipped with a $50 \mathrm{mm}$ focal-length lens and a focal ratio $f_\# = 11$, is used to image the flow with a resolution of approximately $16.9 \mathrm{pixels/mm}$. The field of view is cropped over the region of interest on top of the jets, covering $110\times74\mathrm{mm}^2$ ($0.77\mathrm{m}\leq x \leq 0.88\mathrm{m}$ and $|z|\leq 0.37\mathrm{m}$). DiEthyl-Hexyl-Sebacate particles of $\approx 1 \mu\mathrm{m}$ size are used to seed the flow. Illumination is provided by a dual cavity Nd:Yag Quantel Evergreen laser ($200\mathrm{mJ/pulse}$ at $15\mathrm{Hz}$) and a set of cylindrical and spherical lenses.

Flow statistics are computed from an ensemble of $2000$ image pairs acquired at a sampling frequency of $15\mathrm{Hz}$. Image pre-processing to remove the background is based on the eigenbackground removal procedure \citep{Mendez2017pod-piv}. The PIV processing software implemented by the Experimental Thermo-Fluid-Dynamics Group of the University of Naples Federico II is used. A multi-pass cross-correlation algorithm \citep{soria1996piv} with window deformation \citep{scarano2001iterativeimgdef} is applied to the sequence of images. The software includes advanced interpolation schemes and weighting windows to improve the spatial resolution and precision of the process \citep{Astarita2006PIV,Astarita2007PIV}. The PIV interrogation process had a final interrogation window size of $48\times48$ $\mathrm{pixels}^2$ and 75\% of overlap, resulting in $1.4\mathrm{vector/mm}$. The expected uncertainty of the displacement field is approximately $0.1$ pixels \citep{raffel2018piv}. A small time separation ($20\mu s$) between the laser impulses was set to improve the signal-to-noise ratio in the correlation process, accounting for the low seeding density (the jet was not seeded) and large velocity gradients. This resulted in typical displacements of $4$ pixels, leading to an uncertainty of about 2.5\% on the instantaneous velocity fields.

\section{Characterisation of the unperturbed and steady-jet flow}\label{s:Characterisation}
The unperturbed TBL is characterised using planar-PIV measurements in the wall-normal, flow-parallel plane $z = 0$. The same experimental setup as for the optimisation experiments is used to ensure that the flow characterisation is representative of the actual flow conditions during the learning process. The field of view of the PIV snapshots expands along the slot-jet length in the streamwise direction and several boundary layer thicknesses away from the wall. The PIV apparatus is the same as described in section~\ref{ss:PIV}, modifying the optical path and the orientation of the laser and cameras. Statistics of the flow field are computed using Ensemble Particle Tracking Velocimetry \citep{cowen1997hybrid}, with the polynomial fitting of the particle clouds to improve the spatial resolution \citep{aguera2016EPTV}. The final averaging bin is $200\times1~\mathrm{pixels}$ in a total of $2000$ images.

\begin{figure}
    \centering
    \includegraphics[width=0.9\linewidth]{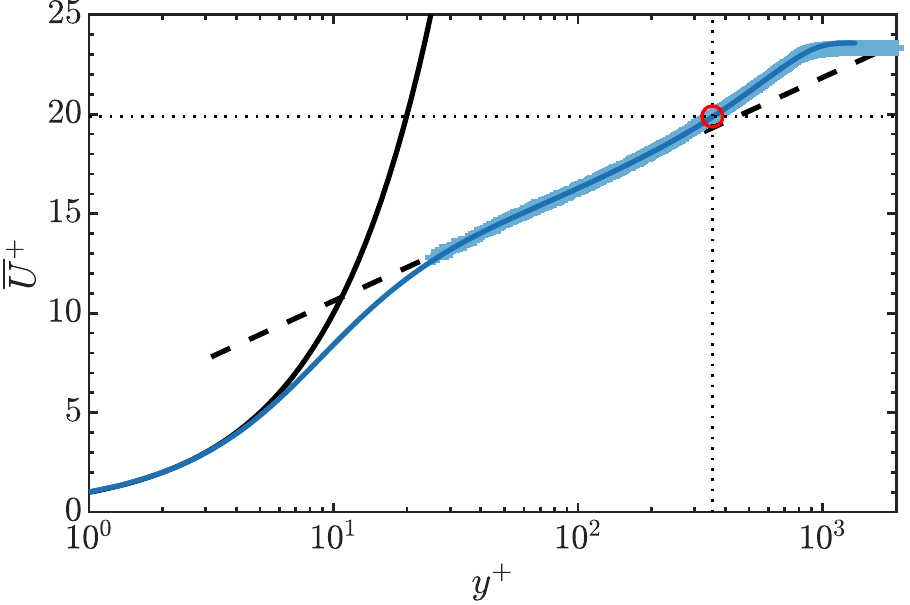}
    \caption{Profile of the unperturbed TBL at the slot-jet streamwise location. PIV experimental data \sy{BlueTBL1}{+} is depicted together with its fit based on \citet{chauhan2009} \lcap{-}{BlueTBL2}. The viscous sublayer \lcap{-}{black} $\overline{U}^+ = y^+$, and the logarithmic law \lcap{--}{black} $\overline{U}^+=\frac{1}{0.41} \mathrm{ln}y^+ + 5.0$ are included for reference. The plane $y^+\approx 350$ ($\overline{U}^+\approx 20$ and $y=10.6\mathrm{mm}$, $\overline{U}=10.3\mathrm{m/s}$) is the location where wall-parallel, planar PIV is performed \sy{red}{o}.}
    \label{fig:TBLprofile}
\end{figure}

The TBL parameters are calculated by fitting the experimental data on the composite profile proposed by \citet{chauhan2009}. The mean streamwise velocity profile, $\overline{U}^+$, is depicted in figure~\ref{fig:TBLprofile} together with the fitted data. The boundary layer profile satisfactorily agrees with the law of the wall. The TBL parameters are attached in table \ref{tab:TBL}, including the friction-based Reynolds number $\mathrm{Re}_\tau$, the Reynolds number based on momentum thickness $\mathrm{Re}_\theta$, the displacement and momentum thickness $\delta^*$ and $\theta$, the shape factor $H_{12}=\delta^*/\theta$ and the friction velocity $u_\tau$. The boundary layer thickness $\delta$  is approximated as $\delta \approx \delta_{99}$, such that $U(y=\delta_{99})=0.99U_\infty$. An uncertainty quantification on the estimation of the TBL parameters is carried out based on the analysis tool by \citet{Castellanos2021PIVuncertainty} for PIV/EPTV measurements. The uncertainty of the displacement thickness $\delta^*$ and boundary layer thickness $\delta$ are found to be below $\pm 2\%$, while for the freestream and local friction velocity, the error lies below $1\%$. The position of the wall is estimated based on the experimental measurements and from the corrections during the fitting process with an error below $\Delta y^+ = \pm 5$. 

\begin{table}[t] 
    \centering
    \caption{Parameters of the unperturbed TBL at the slot-jet streamwise location.}
    \begin{tabular}{>{\centering}p{0.125\linewidth-2\tabcolsep}
            >{\centering}p{0.125\linewidth-2\tabcolsep}
            >{\centering}p{0.125\linewidth-2\tabcolsep}
            >{\centering}p{0.125\linewidth-2\tabcolsep}
            >{\centering}p{0.125\linewidth-2\tabcolsep}
            >{\centering}p{0.125\linewidth-2\tabcolsep}
            >{\centering}p{0.125\linewidth-2\tabcolsep}
            >{\centering\arraybackslash}p{0.125\linewidth-2\tabcolsep}}
    \toprule
    $\mathrm{Re}_\tau$ & $\mathrm{Re}_\theta$ & $H_{12}$ & $\delta^*/\delta_{99} $ & $\theta/\delta_{99}$ & $\delta_{99}$ [mm] & $u_\tau$ [m/s] & $U_\infty$ [m/s]\\
    \midrule
    876 & 2186 & 1.43 & 0.153 & 0.107 & 26.3 & 0.52 & 12.1\\ 
    \bottomrule
    \end{tabular}
    \label{tab:TBL}
\end{table}

The diagnostic-plot method proposed by \citet{sanmiguel2017diagnostic} is used to determine whether the TBL can be considered well-behaved from the experimental PIV data for the streamwise velocity $\overline{U}$ and the fluctuations $(\overline{u^2})^{1/2}$ in the wake region. The main advantage of this technique is that it only requires measurements in the outer layer, which can be obtained with high accuracy with PIV/EPTV. The diagnostic-plot parameters $\alpha$ and $\beta$ are reported in table \ref{tab:diagnostic}, showing good agreement with numerical simulation from \citet{jimenez2010} for a similar $\mathrm{Re}_\theta$.

\begin{table} 
\caption{Diagnostic-plot parameters of the unperturbed TBL at $x=x_\mathrm{jet}$ and the DNS data from \citet{jimenez2010}. Error with respect to the method by \citet{sanmiguel2017diagnostic} between brackets.}\label{tab:diagnostic}
\begin{tabular*}{\tblwidth}{@{}LLCC@{}}
\toprule
Case & $\mathrm{Re}_\theta$ & $\alpha$ & $\beta$ \\ \midrule
Unperturbed TBL & 2186 & 0.305 ($+2.2\%$) & 0.270 ($+3.5\%$) \\
\citet{jimenez2010} & 1968 & 0.287 ($-1.0\%$) & 0.254 ($+0.3\%$) \\
\bottomrule
\end{tabular*}
\end{table}

Two reference flow conditions are assumed to evaluate the performance of the controller in terms of Nusselt number $\mathrm{Nu}$: the unperturbed TBL, denoted by the subscript `$0$'; and the steady-jet condition characterised by a steady operation of all the slot jets, identified by subscript `$sj$'. 

\begin{figure*}[]
\centering
    \includegraphics*[width=0.95\linewidth]{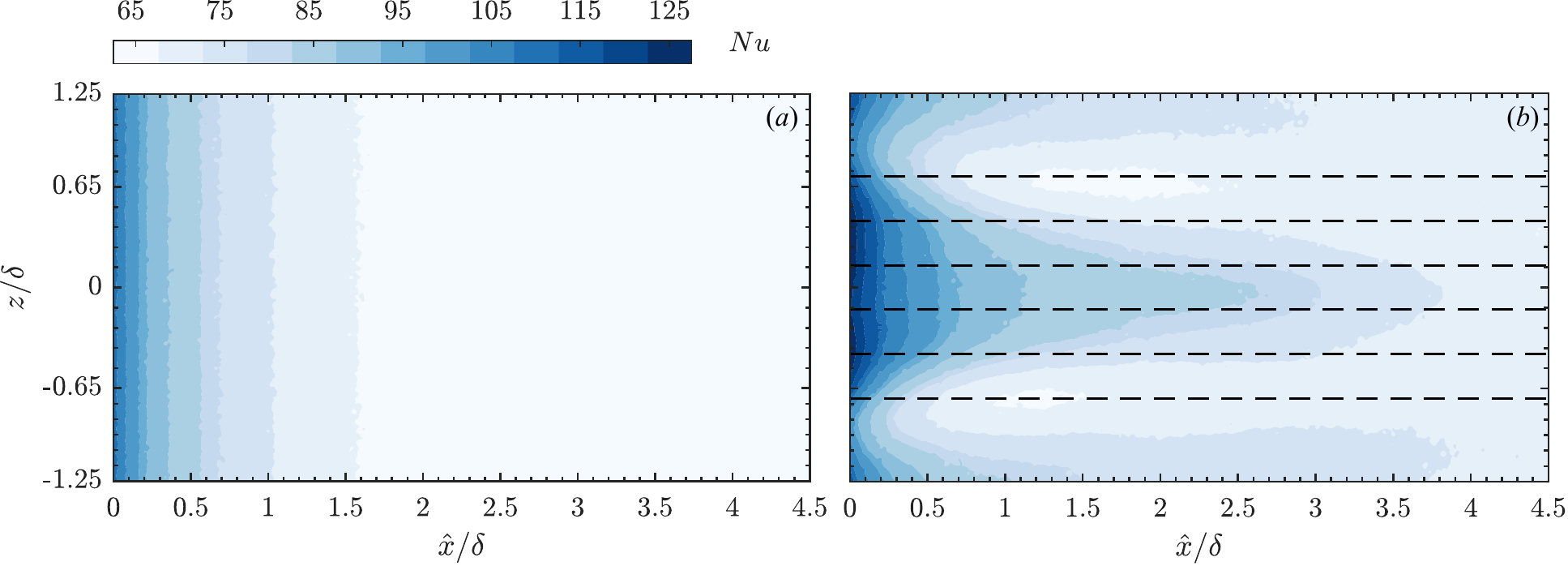}
\caption{\label{fig:reference_Nu} Nusselt number distribution downstream of the slot jet for the unperturbed TBL (a) and the steady-jet actuation (b). The spanwise positions of the slots are highlighted \lcap{--}{black} for reference.} 
\end{figure*}

Infrared thermography measurements are performed directly downstream of the slot jets in a region comprising $4.5\delta \times 2.5\delta$. The spatial distributions for both $\mathrm{Nu}_\mathrm{0}$ and $\mathrm{Nu}_\mathrm{sj}$ are shown in figure~\ref{fig:reference_Nu}. The Nusselt number distribution for the unperturbed case agrees with the literature, following a power-law decay $\mathrm{Nu} \sim x^{-0.2}$ as expected for a conventional TBL \citep{Lienhard2020}.

\begin{table}[t] 
    \centering
    \caption{Reference performance indicators of the steady jet (per jet) and unperturbed TBL. Average quantities computed over the depicted area in figure~\ref{fig:reference_Nu}.}
    \label{tab:steady-jet}
        \begin{tabular}{>{\centering}p{0.2\linewidth-2\tabcolsep}
                >{\centering}p{0.245\linewidth-2\tabcolsep}
                >{\centering}p{0.185\linewidth-2\tabcolsep}
                >{\centering}p{0.185\linewidth-2\tabcolsep}
                >{\centering\arraybackslash}p{0.185\linewidth-2\tabcolsep}}
        \toprule
        $\dot{m}_{\mathrm{sj}}$ [g/s] & $U_\mathrm{jet,sj}$ [m/s] & $\overline{\mathrm{Nu}}_{0}$ & $\overline{\mathrm{Nu}}_{\mathrm{sj}}$ & $\overline{E}$ \\ \midrule
        5.29 & 7.56 & 67.2 & 77.3 & 0.15\\
        \bottomrule
        \end{tabular}
\end{table}

A summary of the actuation parameters for the steady-jet condition is reported in table~\ref{tab:steady-jet}. The combination of the 6 slot jets provides a mass injection of $\dot{m}_{\mathrm{sj}} = 5.29 \mathrm{g/s}$. The steady-jet actuation thus features a bulk velocity of $U_\mathrm{jet,sj} = 7.56 \mathrm{m/s}$. This corresponds to a velocity ratio of $R = 0.62$ with respect to the freestream velocity. It is worth remarking that, since the experiments are performed at constant pressure conditions, the velocity ratio is fixed for all the experiments of this work. Such circumstances are not valid under other actuation conditions as is the case of constant mass or volumetric flow rate. 

The Nusselt number distribution exhibits a symmetric pattern with high values of $\mathrm{Nu}$ in the vicinity of the air injection. This pattern resembles that of the spanwise-aligned slot jet tested by \citet{Castellanos2022slotjet}, although lower values of $\mathrm{Nu}$ are achieved. A worth-noting feature of the $\mathrm{Nu}$ distribution along the spanwise direction is its local reduction downstream of the slot jets at both ends, i.e. the left- and rightmost actuators. This feature suggests a local reduction of streamwise velocity promoted by the diversion of fluid towards the centre line of the actuator and away from the wall. The wall-normal injection of fluid in the lower region of the boundary layer is locally improving the mixing downstream; however, a partial flow blockage may occur. The flow is diverted out of the wall by the jets and, secondarily, the flow is pushed sideways, surrounding the actuator area. This secondary effect might explain the spanwise modulation in $\mathrm{Nu}$, in which a sudden increase of $\mathrm{Nu}$ is observed for $|z/\delta| \gtrapprox 0.75$.

The effect of the jet is persistent even $4\delta$ downstream of the actuator; nonetheless, the increment of $\mathrm{Nu}$ is very localised in the spanwise direction, mainly affecting the area downstream of the jets. Consequently, most of the quantities regarding heat transfer analysis will be spatially averaged, either over a preferred spatial domain (namely over-lined symbols, e.g. $\overline{\mathrm{Nu}}$) or along the streamwise or spanwise directions ($\langle \mathrm{Nu} \rangle_x$ and $\langle \mathrm{Nu} \rangle_z$, respectively). In the following, the average Nusselt number is defined as the integral average across the spatial domain $D = \{(\hat{x},z) \in \mathbb{R}^2:  0\leq \hat{x}/\delta \leq 4.5; -1.25 \leq z/\delta \leq 1.25\}$ with measure $m(D)$, i.e.
\begin{equation} \label{eq:Num}
    \overline{\mathrm{Nu}} = \frac{1}{m(D)}\iint_{D}^{} \mathrm{Nu}(\hat{x},z) \,d\hat{x}\,dz
\end{equation}

Convective heat transfer enhancement is defined as the favourable increment of heat flux at the wall with respect to a reference condition. The enhancement $E$ for a given control action is quantified based on the definition proposed by \citet{Castellanos2022slotjet}, which is the relative change in Nusselt number with respect to the unperturbed TBL, namely $\overline{\mathrm{Nu}}_\mathrm{0}$,
\begin{equation} \label{eq:E}
    E = \frac{\overline{\mathrm{Nu}}}{\overline{\mathrm{Nu}}_\mathrm{0}}-1.
\end{equation}

A summary of the main features of the unperturbed case and steady-jet actuation is included in table~\ref{tab:steady-jet}.

%
\section{Machine Learning Control}\label{s:MLC}
The considered optimisation framework in this work lies under the definition of Machine Learning Control \citep{BruntonNoack2015review,duriez2017book}. The outstanding performance of MLC is reported in several flow control problems such as jet mixing enhancement \citep{Wu2018jet}, stabilisation of the fluidic pinball \citep{cornejo2021gMLC} or drag reduction of shedding flows \citep{Castellanos2022LGPCvsRL}, among others. In this section, we formulate the optimisation problems and discuss the caveats of the definition of the cost function. The adopted genetic algorithm is described, and the training process is outlined.


\subsection{Formulation of the optimisation problem}\label{ss:opt_problem}

The control optimisation is posed as a regression problem of the second kind. In the following, the actuation command is denoted by $b$. The optimised control is the strategy that minimises the cost function with a control law $\bm{b}(t) = {K}(t;\bm{\Xi})$, where $\bm{b}(t) = (b_1, \ldots ,b_{N_b})^T$ comprises $N_b$ actuation commands and $\bm{\Xi} = (\Xi_1, \ldots , \Xi_{N_p})^T$ consist of $N_p$ control parameters. In this study, the control authority is driven by two sets of jets ($N_b = 2$), depending on $N_p = 5$ control parameters as later defined. The optimisation problem is equivalent to finding ${K}^{*}$ such that,
\begin{equation}\label{eq:Optproblem}
    \begin{aligned}
        {K}^{*}(\bm{\Xi}) = \underset{{K \in \mathbb{K}}}{\arg\min}~~ J({K}(\bm{\Xi}))
    \end{aligned}
\end{equation}

with $\mathbb{K}: \mathcal{P} \mapsto \mathcal{B}$ being the space of control laws. For this problem, $\mathcal{P}$ is the control parameter space and $\mathcal{B}$ is the output space gathering all the possible control laws. In other words, the resultant controller maps $N_p$ control parameters into $N_b$ actuation commands. In general, the model-free optimisation of control laws (Eq.~\eqref{eq:Optproblem}) is a challenging non-convex regression problem of the second kind. For a given argument (control parameters, $\bm{\Xi}$) the optimal action is commonly unknown, as is the case for most of the experimental investigation. Genetic algorithms are employed to tackle this problem, as described in the following sections.

The set of 6 slot jets is divided into two groups (see figure~\ref{fig:SetUp}e). Slots jets are numbered from one spanwise end to the other, and assigned independent control laws, $b_1$ and $b_2$ for odd and even jets, respectively. The actuation is the pulsation of a set of jets whose average mass flow rate scales with the $DC$. Periodic forcing with frequencies $f_1$ and $f_2$ is considered. We also investigate the relevance of the phase shift $\phi$ between the odd and even actuators. Thus, a total of five control parameters are to be optimised,
\begin{equation}\label{eq:params}
    \bm{\Xi} = [f_1,f_2,DC_1,DC_2,\phi]
\end{equation}

It is to be noted that the phase $\phi$ becomes irrelevant when the actuation frequency of both sets of actuators is not the same or a multiple between them. This is the reason why most of the flow-control studies based on pulsed actuators discard the phase as a control parameter \citep[e.g.][]{Wu2018jet,minelli2020lgac}. However, based on our previous finding on a similar experimental investigation \citep{Castellanos2022slotjet}, it is expected to find a characteristic frequency at which heat transfer enhancement is maximised independently of the duty cycle. The existence of a preferred frequency could drive the controller towards the same frequency for all the actuators, making the phase a relevant control parameter.

The frequency $f$ of the pulsed jets is also treated in its dimensionless form, the Strouhal number $\mathrm{St}$. The definition of $\mathrm{St}$ proposed by \citet{Castellanos2022slotjet} is adopted, which reads
\begin{equation}\label{eq:St}
    St = \frac{f \cdot \delta}{c},
\end{equation}
being $c \approx 10 u_\tau = 5.2 \mathrm{m/s}$ the convection velocity of coherent structures in the near-wall region, whose value is estimated based on literature \citep{jimenez2004lsturb}. Hence, according to this definition, the frequency corresponding to $\mathrm{St} = 1$ is the one exciting the turbulent coherent motions with a size of the order of $\delta$ that move within the TBL with a velocity $c$.

\citet{Fan2017jetcontrol} reports the performance of periodic forcing for the mixing enhancement of a turbulent jet using linear genetic programming control. \citet{Wu2018jet} extends the previous work to multi-frequency forcing although the algorithm converges to a pure periodic forcing. Despite the potential advantage of a more complex actuation command derived from a multi-frequency forcing, a single-frequency periodic forcing is preferred because of its ease of implementation and the interpretability of the resultant control law. The corresponding open-loop control laws read,
\begin{align}\label{eq:control}
    \begin{split}
        b_1(t) & = H\left( \sin(2\pi f_1 t) + \kappa(DC_1)\right) \\
        b_2(t) & = H\left( \sin(2\pi f_2 t + \phi) + \kappa(DC_2)\right)
    \end{split}
\end{align}
being $H()$ the Heaviside function and $\kappa \in [-1, 1]$ a proportional scaling of the duty cycle $DC$. With finite data acquisition frequency, only discrete frequencies, duty cycles and phases are possible. Thus, the control parameters are limited to the following ranges: $f_1,f_2 \in [30,285] \mathrm{Hz}$, $DC_1,DC_2 \in [25\%,76\%]$, and $\phi \in [0, \pi]$. These ranges represent the search space in which the optimisation algorithm seeks the optimal control law. The control law $b(t)$ takes binary values 1 and 0 depending on whether the slot jet is ON or OFF, i.e. blowing or not. The apparatus constraints do not allow fine control of the mass flow rate at a specific instant of time, limiting the control to a fully-modulated actuation. Nonetheless, it is worth noting that a simple periodic forcing may modify the dominant frequency of the phenomena within the TBL, or even alter the broadband frequency content, exploiting nonlinear frequency cross-talk \citep{BruntonNoack2015review}.

\subsection{Cost function} \label{ss:cost_function}
The objective function for training should include both the heat transfer rate and the cost of the actuation. \citet{manglik2003heat} describes a performance evaluation criteria (PEC) for heat exchangers relating the change in heat transfer coefficient $h$ with the required pumping power $P$. This classical interpretation for heat transfer enhancement reads,
\begin{equation}
    \frac{hA/h_{0} A_{0}}{(P/P_{0})^{1/3}(A/A_{0})^{2/3}} = \frac{\mathrm{Nu}/\mathrm{Nu}_{0}}{(f/f_{0})^{1/3}},
\end{equation}
where $A$ is the heat transfer surface area, $f$ is the skin-friction coefficient, and the subscript `$0$' refers to the unperturbed condition. In this work, the PEC is adapted so that the cost function also accounts for the enhancement in heat transfer and the required power. Regarding the former, a direct figure of merit of the heat transfer enhancement $E$ (see Eq.~\eqref{eq:E}) is used. The actuation cost is estimated as the jet power $P_{j} = \dot{m}\frac{1}{2}\rho U_\mathrm{jet}^2$. This is obtained assuming incompressible flow (i.e. $\rho$ is constant) and monitoring the value of the overall mass flow rate through the slots. Hence, the objective function is divided into $J_a$, quantifying $E$, and $J_b$, related to the jet power required by the control law. The cost $J_a$ is the main target of this investigation and is defined as
\begin{equation}\label{eq:Ja}
    J_a = 1 - E = 1 - \left( \frac{\overline{\mathrm{Nu}}}{\overline{\mathrm{Nu}}_\mathrm{0}} - 1 \right)
\end{equation}
in which the Nusselt number distribution is averaged on a spatial domain of interest as described in Eq.~\eqref{eq:Num}. The $\mathrm{Nu}$ values are obtained as the ensemble average of the snapshots on the whole duration of the experiment of each tested individual, as described in \S\ref{ss:training_standard}.

The cost $J_b$ is conceived as a penalisation term to quantify the power requirement of the controller. Since the pressure difference between the line and the outflow is practically fixed by the value set by the pressure-relief valve, the power requirements scale quite well with the mass flow rate. Thus, $J_b$ is defined as the ratio of the mass flow rate of air injected by the actuators to the steady-jet actuation,
\begin{equation}\label{eq:Jb}
    J_b = \frac{\dot{m}}{\dot{m}_\mathrm{sj}}
\end{equation}

The definition of $J_b$ is analogous to an average duty cycle after combining $b_1$ and $b_2$. The inclusion of the penalisation term $J_b$ in the cost function $J$ is of paramount importance for a problem of this type, in which the target defined by $J_a$ is strongly affected by the power consumption. The penalisation term $J_b$ is a direct figure of merit of the mass flow injection, and it also affects $J_a$, since higher mass flow rates imply higher heat transfer rates. The resultant cost function reads as
\begin{equation}\label{eq:J}
    J = J_a + \gamma J_b
\end{equation}

being $\gamma$ an arbitrary penalisation coefficient to weigh the relevance of $J_b$ in the optimisation process. This cost function assumes $J_a$ and $J_b$ to have the same order of magnitude. Furthermore, their rate of change should be similar when passing from one control law to another so that the algorithm does not tend to minimise the cost function by only focusing on the most relevant term of $J$. The proposed costs $J_a$ and $J_b$ are both of order unity and their gradient is very similar.

\begin{figure}[]
\centering
    \includegraphics*[width=0.95\linewidth]{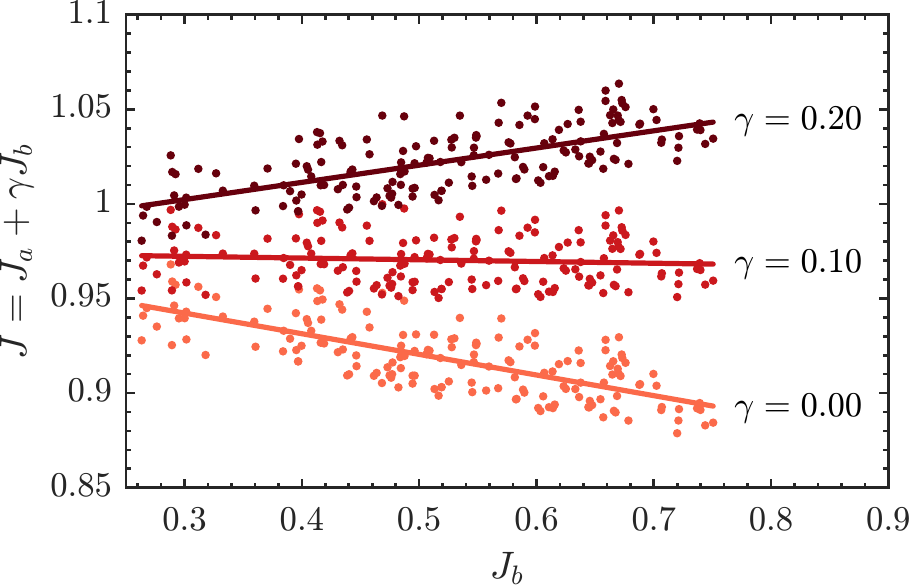}
\caption{\label{fig:alpha} Preliminary analysis of the cost function performed on $300$ parameter combinations from a random sampling. Dependency of $J=J_a+\gamma J_b$ on $J_b$ for different values of the penalisation coefficient $\gamma$. The dots are coloured from bright to dark for increasing value of $\gamma$.} 
\end{figure}

In order to avoid the optimisation to reach a trivial optimal solution (e.g. either focusing only on minimising the energy input, or maximising only heat transfer, but not weighing both at the same time) we need to find a fair weight for the penalisation term in the optimisation process. A large value would lead the optimisation to focus only on minimising energy input, while a small value would lead to focus only on maximising heat transfer, regardless of the energy input. In our empirical analysis, we aim for a value of $\gamma$ which balances these two tendencies, i.e. with a cost function $J$ which is not driven by $J_b$ alone.
The penalisation coefficient $\gamma$ is determined by a parametric study. The cost $J$ is evaluated for $300$ randomly sampled individuals so that the dependency of $J_b$ on $J$ is evaluated, as illustrated in figure~\ref{fig:alpha}. By setting $\gamma = 0.1$, the bias towards large duty cycles for a maximisation of the convective heat transfer is avoided, as well as the tendency towards minimising solely the cost of actuation by reducing $J_b$.

\subsection{Linear Genetic Algorithms as regression solver} \label{ss:lgac}

\begin{figure*}
    \centering
    \includegraphics[width = 0.99\linewidth]{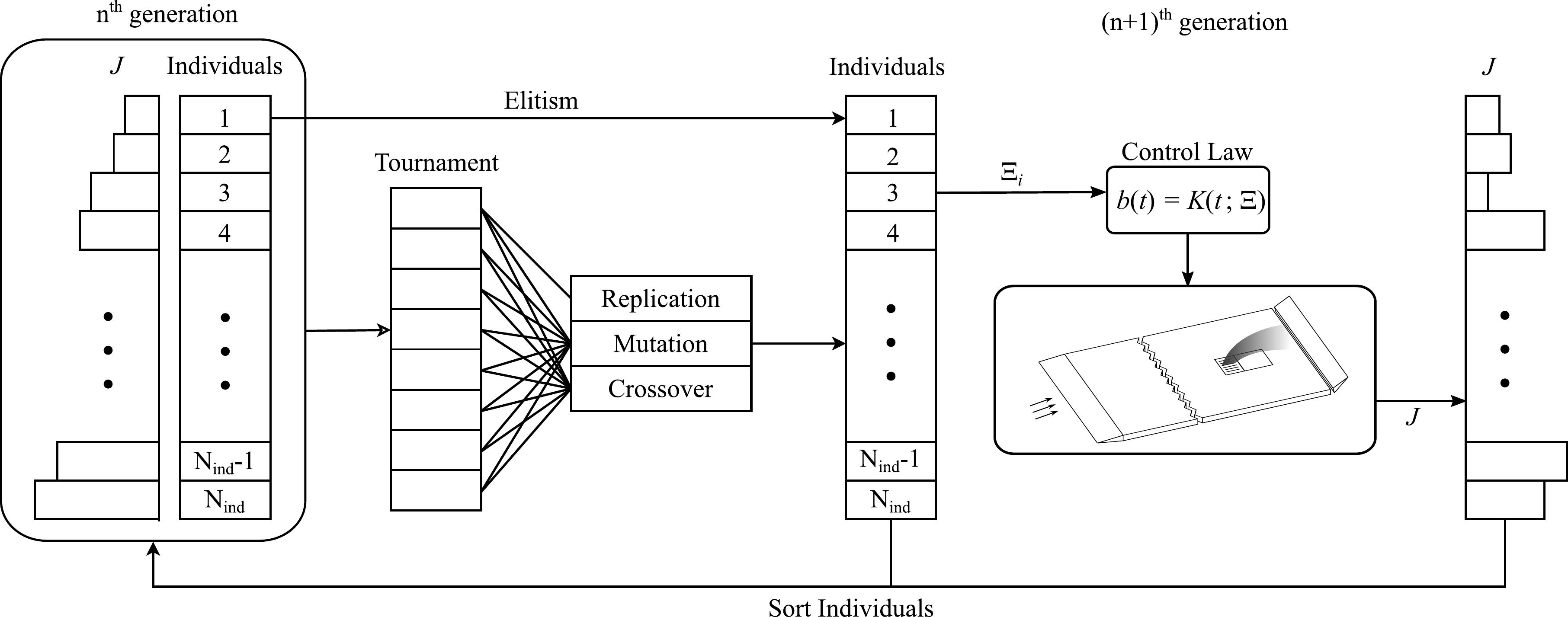}
    \caption{Open-loop Linear Genetic Algorithm Control workflow.}
    \label{fig:flowchart}
\end{figure*}

The GA used in this study, namely Linear Genetic Algorithm Control (LGAC), is based on a binary encoding scheme first proposed by \citet{holland1992adaptation} and on the implementation suggested by \citet{Wahde2008book}. This evolutionary algorithm applies biological-inspired operations to select the fittest control laws, which are also referred to as \textit{individuals} to comply with the evolutionary terminology. LGAC allows a simple implementation of the genetic operators for Multiple-Input-Multiple-Output control. A similar algorithm is employed by \citet{minelli2020lgac} to control the wake of a bluff body.

The authors are aware of the wide variety of optimisation methods, either based on classical optimisation theory, such as downhill simplex algorithms, or machine-learning-based strategies such as deep reinforcement learning. However, LGAC is preferred for its simplicity in its implementation in an experimental investigation. Further work is being developed to compare LGAC with alternative tools in a similar experimental plant.

The open-loop learning process is sketched in figure~\ref{fig:flowchart}. An individual is a set of parameters in Eq.~\eqref{eq:params}, which describe a periodic forcing, i.e. the mass flow injection of the jets based on Eq.~\eqref{eq:control}. For the first generation, a random Monte-Carlo sampling generates a population of individuals to explore the solution space. The control law $b(t)$ of each individual is fixed during the evaluation. The performance of the individuals is quantified by its cost $J$ as in Eq.\eqref{eq:J}. Once the entire population is evaluated, the most performing individuals are selected with the tournament selection method and combined based on the genetic operators (crossover, mutation, replication) to generate the following population.

Elitism operation is first invoked to ensure no degradation in the performance of the following generation by directly evolving the best individuals. The crossover operation exploits the learned data by recombining well-performing individuals. Two individuals randomly exchange their instructions to generate a new pair of individuals. On the other hand, exploration is accomplished by the mutation operation, which randomly alters part of an individual. Replication is a memory operator since it assures that good structures are not lost in the evolution process. The operator is chosen based on their corresponding probabilities, which are user-defined parameters that can be tuned to strengthen either the exploration or exploitation nature of the algorithm. The tournament size and the genetic operators' probability are chosen following the recommendations by \citet{duriez2017book} and \citet{Castellanos2022LGPCvsRL} and summarised in table~\ref{tab:GAparameters}.

\begin{table}[h]
    \centering
    \caption{Selection of LGAC parameters} \label{tab:GAparameters}
    \begin{tabular}{lc}
        \toprule
        Number of controllers      & 2 \\
        Population size            & 100 \\
        Number of generations      & 10 \\ 
        Bits per variable          & 8 \\
        Tournament selection size  & 7   \\ 
        Crossover probability      & 0.6 \\ 
        Mutation probability       & 0.3 \\
        Replication probability    & 0.1 \\
        Elitism                    & 1   \\ 
        $f_1$ and $f_2$ range      & 30 - 285 Hz   \\
        $DC_1$ and $DC_2$ range    & 25 - 76 \%   \\ 
        $\phi$ range               & $0 - \pi$ \\ \bottomrule
    \end{tabular}
\end{table}

The genetic evolution process is repeated for every generation until reaching the stopping criterion. A fixed number of generations $N_{g} = 10$ is defined based on previous applications of evolutionary algorithms \citep{Yu2021GA_drag_slot_jets,Wu2018jet,cornejo2021gMLC}.

\subsection{Training Process} \label{ss:training_standard}

The training process is designed to comply with several experimental and time constraints. The LGAC optimisation process features a pool size of $100$ individuals per generation. The best individual of the $10^{th}$ generation is the final optimised control law. The main constraint in terms of training time is the data acquisition and evaluation, while the LGAC optimisation requires a negligible fraction of time compared to the experiment itself. The training process illustrated in figure~\ref{fig:flowchart} is divided into two main blocks: measurement and evaluation. 

The measurement block is directly related to the experimental data acquisition. The experiment is performed for each non-repeated individual within a given generation. Once the data is acquired for a given population, the process enters into the evaluation block in which the Nusselt number $\mathrm{Nu}$ and the costs $J_a$, $J_b$ and $J$ are computed for each individual. 

The existence of outliers when evaluating a population may bias the optimisation process. To guarantee the repeatability of the experiment while avoiding potential outliers, it was decided to perform each measurement and evaluation of individuals twice. If the error between the costs from the two independent repetitions lies below the experimental uncertainty ($5\%$), then the mean value of the cost $J$ of the two runs is considered. Otherwise, the measurement and evaluation of the detected outliers are repeated a third time. If, despite the 3 repetitions, the value of the error is still high, a high-cost value ($J = 10^{36}$) is assigned to discard the individual and prevent it from affecting the LGAC training process. A summary of the measurement and evaluation process is included in algorithm~\ref{alg:rep}.

\begin{algorithm}
\caption{Measurement and Evaluation of a Generation $g$}\label{alg:rep}
\KwData{$\xi = 5\%$, experimental uncertainty}
\KwResult{Unbiased individuals in a generation}

\underline{Measurement and Evaluation of rep. 1 and 2}\;
$r = 1$, repetition count\;
\Repeat{$r \leq 2$}{
    \ForAll{$i \in g$}{
        \textbf{Measure block}: acquire data: $(T_\mathrm{w}, T_\mathrm{aw},\dot{m}, T_\infty, \ldots)$\;
    }
    \ForAll{$i \in g$}{
        \textbf{Evaluation block}: evaluate data and compute costs: ($\mathrm{Nu}, J_a, J_b, J$)\;
    }
    $r+=1$\;
}

\underline{Outliers detection}\;
\ForAll{$i \in g$}{
    \textbf{Error estimation}: $e = (|J_1 - J_2|)/max(J_1,J_2)$\;
    \uIf{$e \leq \xi$}{$J = mean(J_1,J_2)$}
    \Else{Outlier detected: $i\in g^* \subset g$}
}

\underline{Measurement and Evaluation for outliers ($r=3$)}\;
\ForAll{$i \in g^*$}{
    \textbf{Measure block}: acquire data: $(T_\mathrm{w}, T_\mathrm{aw},\dot{m}, T_\infty, \ldots)$\;
}
\ForAll{$i \in g^*$}{
    \textbf{Evaluation block}: evaluate data and compute costs: ($\mathrm{Nu}, J_a, J_b, J$)\;
}
\ForAll{$i \in g^*$}{
    \textbf{Error estimation}: 
    $e_1 = (|J_3 - J_1|)/max(J_3,J_1)$\;
    $e_2 = (|J_3 - J_2|)/max(J_3,J_2)$\;
    \uIf{$e_1 \leq e_2$ \& $e_1 \leq \xi$}{
        $J = mean(J_1,J_3)$\;
    }
    \uElseIf{$e_2 < e_1$ \& $e_2 \leq \xi$}{
        $J = mean(J_2,J_3)$\;
    }
    \Else{
        Assign bad value: $J = 10^{36}$\;
    }
}
\end{algorithm}

The number of outliers is considerably reduced by defining a proper experiment duration that ensures the convergence towards a new steady state upon actuation. The experiment duration for each individual is $t_{ind} = 120\mathrm{s}$ divided into 4 tasks: (1) convergence time for $T_\mathrm{aw}$, $t_1 = 45\mathrm{s}$; (2) data acquisition for $T_\mathrm{aw}$, $t_2 = 20\mathrm{s}$, (3) convergence time for $T_\mathrm{w}$, $t_3 = 30\mathrm{s}$; and (4) data acquisition for $T_\mathrm{w}$, $t_4 = 20\mathrm{s}$. Following the description in \S\ref{ss:IR}, $t_2$ and $t_4$ correspond to the acquisition time for \textit{cold} and \textit{hot} images, respectively. The values of the time intervals are set empirically based on preliminary tests spanning a range of different actuation settings. The quick convergence to temperature differences, comparable to the noise-equivalent temperature difference of the IR camera, is eased by the small thermal inertia of the foil. The acquisition time is derived from the sampling rate of $10$ snapshots per second and a total of $200$ snapshots, which is accurate enough to assure the convergence of the $\overline{\mathrm{Nu}}$ temporal average. On the other hand, the estimation of $t_1$ and $t_3$ is of paramount importance to guarantee the steady-state condition before the data acquisition. The interval $t_1$ is the required time for the heat-flux sensor to converge towards $T_\mathrm{aw}$ when changing from one individual to another, while $t_3$ is the time for a given individual to reach the steady-state $T_\mathrm{w}$ upon the activation of $q''_j$ supply, i.e. turning on the power supply (see \S\ref{ss:IR} for more details). It is worth noting that the flow meter and the thermocouple are synchronised with the IR thermography so that their acquisition is triggered during $t_1$ and $t_3$.

Eventually, the training process is carried out for $1000$ individuals distributed in $10$ generations. Each generation endured around $7$ hours of experimental testing and evaluation. The number of detected outliers after the first and second repetition was of the order of $5-10\%$ of the pool size; however, after the additional repetition of the experiment, the resultant number of unsuccessful evaluations is reduced to $\leq1\%$ per generation. A total of $977GB$ of data are generated during the whole training process. Despite the experimental equipment and the data storage requirements, the training process is a rather simple task with minor CPU and memory requirements that could be controlled from a standard personal computer.

\section{Performance of the controller \label{s:Results}}
The performance of the genetically-optimised controller is analysed. The optimisation process consists of $10$ generations with $100$ individuals per generation. After excluding the repeated individuals, either from replication or elitism operations or by chance after crossover or mutation operations, a total of $925$ independent individuals are evaluated in the wind tunnel, with only $6$ individuals discarded as outliers.

\begin{figure}[t] 
    \centering
    \includegraphics*[width=0.85\linewidth]{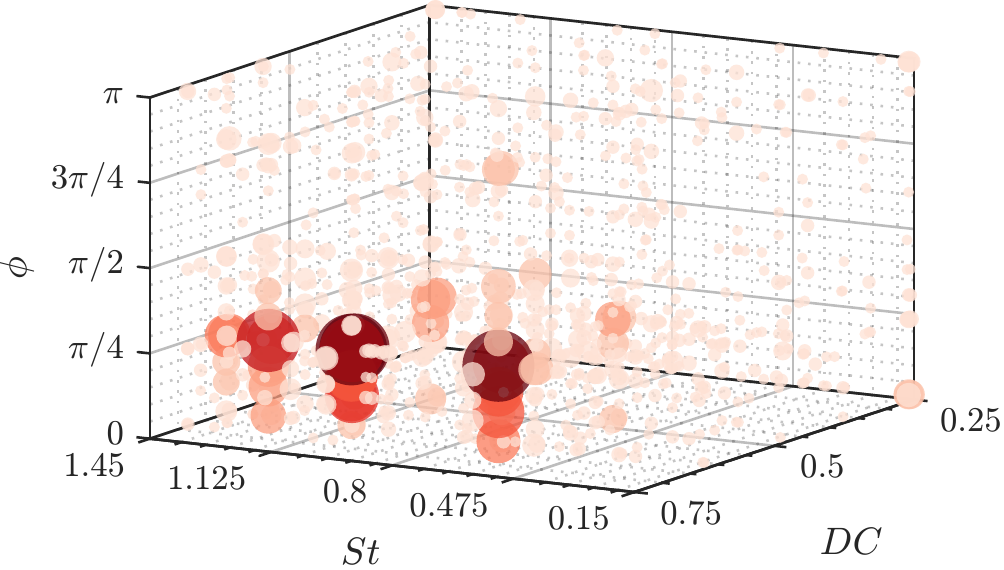}
    \caption{Qualitative representation of the parameter space evaluated by LGAC. The size and colours of the points indicate a greater density of individuals in this area.}
    \label{fig:param_groups}
\end{figure}

The explorative nature of genetic algorithms is satisfied by the mutation of individuals when evolving from one generation to the following. Nonetheless, this kind of algorithm ensures a reasonable exploitation rate directly granted by the replication and elitism and by the cross-over among the best individuals within a generation. The exploration and exploitation capabilities of LGAC in the proposed problem are depicted in figure~\ref{fig:param_groups}, in which each point represents a cluster of individuals and the colour and size of each point define the number of individuals within each cluster. The light, small points show the exploration of the whole feasible solution space; however, the distribution of individuals exhibits a concentration on high $DC$ values, mid-to-high frequencies and low phase differences. These preferred values are denoted by the superscript `\small{$*$}'. 

\normalsize There are three predominant frequencies at which LGAC seeks for the optima, being $f^* = 138, 216~\text{and}~261\mathrm{Hz}$ corresponding to a Strouhal number of $\mathrm{St}^* = 0.70, 1.09~\text{and}~1.32$, respectively. Similarly, the phase between control laws 1 and 2 converges to four possibilities, $\phi^* \approx 9^\circ,18^\circ,29^\circ~\text{and}~43^\circ$. Conversely, the duty cycle swiftly converges to a specific candidate $DC^* \approx 62\%$ for both control laws.

\begin{table*}[] 
    \centering
    \caption{\label{t:best_ind}Evolution of the learning process. Control parameters $\bm{\Xi}$ and costs for the best individual after each generation. The costs are normalised with the values corresponding to the steady-jet actuation $J_0 = 0.9522$, $J_{a,0} = 0.8521$ and $J_{b,0} = 1$.}
    \begin{tabular}{lllllllll}
        \toprule
        Generation & $\mathrm{St}_1$ & $\mathrm{St}_2$ & $DC_1$ (\%) & $DC_2$ (\%) & $\phi$ & $J/J_0$ & $J_a/J_{a,0}$ & $J_b/J_{b,0}$ \\ \midrule
        1,2 & 0.61 & 1.10 & 49.0 & 46.6 & 0.9599 & 0.9504 & 1.0053 & 0.4837 \\
        3,4 & 1.32 & 0.70 & 62.8 & 72.2 & 0.4538 & 0.9394 & 0.9708 & 0.6729 \\
        5-8 & 1.09 & 1.09 & 62.6 & 62.2 & 0.7679 & 0.8861 & 0.9175 & 0.6190 \\
        9   & 1.09 & 1.09 & 61.0 & 62.2 & 0.5061 & 0.8666 & 0.8966 & 0.6119 \\
        \textbf{10} & \textbf{1.09} & \textbf{1.09} & \textbf{62.6} & \textbf{62.2} & \textbf{0.5236} & \textbf{0.8663} & \textbf{0.8951} & \textbf{0.6212} \\ \bottomrule
    \end{tabular}
\end{table*}

The evolution of the best individual after each generation is quantified in table~\ref{t:best_ind}, in which the sets of control parameters are specified together with the achieved costs $J$, $J_a$ and $J_b$. The cost function is normalised with the values corresponding to a steady-jet actuation, i.e. $J_0 = 0.9522$, $J_{a,0} = 0.8521$ and $J_{b,0} = 1$, respectively. From now on, the superscript `$\star$' is used to identify the best individual of a given generation. 

\begin{figure*}[t] 
    \centering
    \includegraphics*[width=0.95\linewidth]{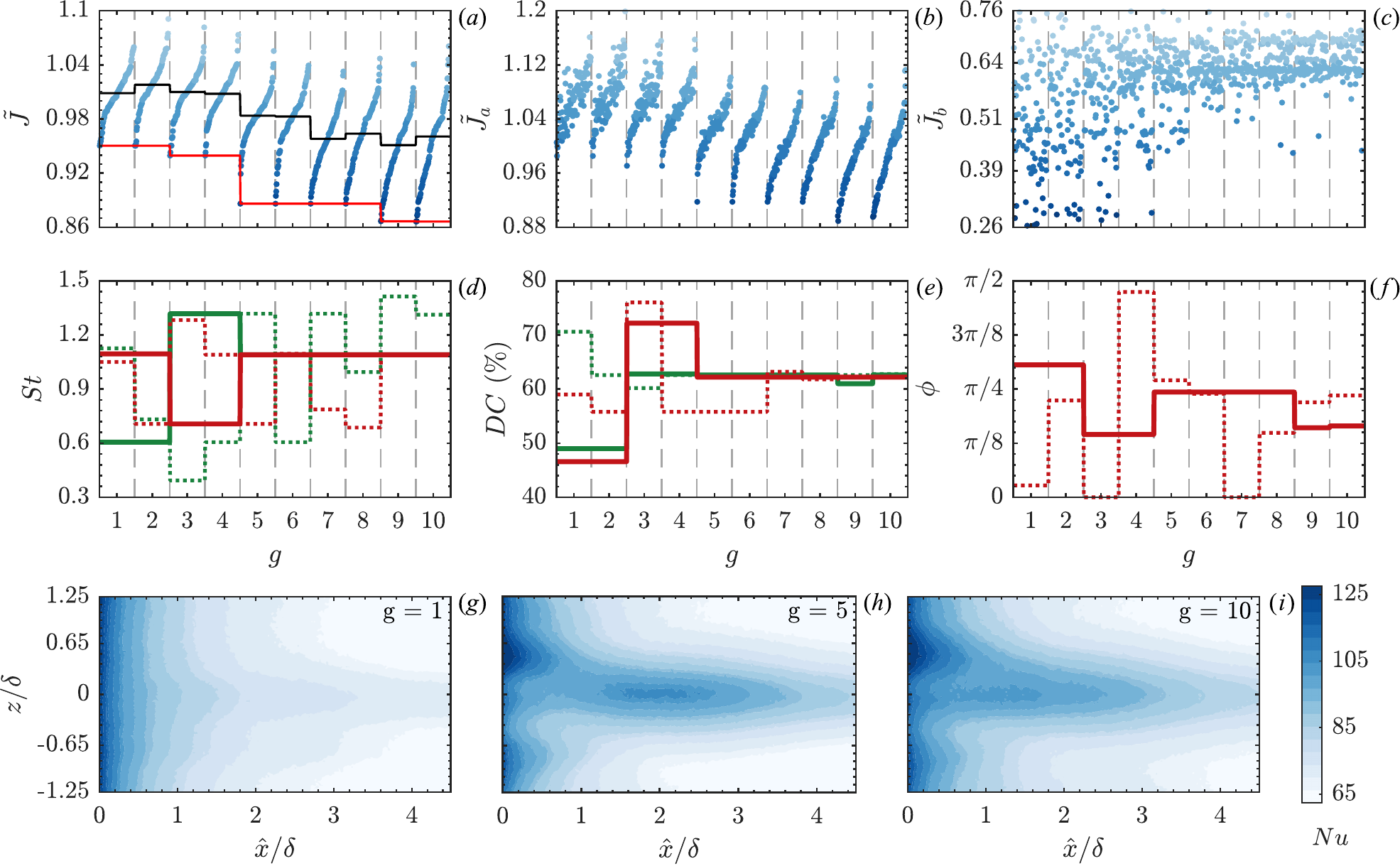}
    \caption{LGAC training process. Each dot represents the costs of each individual normalised with the values corresponding to the steady-jet actuation: $\tilde{J} = J/J_\mathrm{0}$ (a), $\tilde{J_a} = J_a/J_{a,0}$ (b), and $\tilde{J_b} = J_b/J_{b,0}$ (c). The evolution of the best \lcap{-}{red} and median \lcap{-}{black} individuals is included in (a).
    The evolution of the best \lcap{-}{black} and median \lcap{:}{black} set of actuation parameters $\mathrm{St}$ (d), $DC$ (e)  and $\phi$ (f) for the control law $b_1$ \sy{green_lgac}{s*} and $b_2$ \sy{red_lgac}{s*}. 
    The Nusselt number distribution for the best individual of generations 1 (g), 5 (h) and 10 (i).}
    \label{fig:LGAC_progress}
\end{figure*}

Figure~\ref{fig:LGAC_progress} illustrates the evolution of the normalised cost $\tilde{J}=J/J_0$ during the learning process for $10$ generations. Each point represents an individual within a given generation $g$, being sorted by its cost value. After the first generation, a considerable improvement is already achieved with the individual $\bm{\Xi}^\star_1 = [120\mathrm{Hz}, 217\mathrm{Hz}, 49\%, 47\%, 55^\circ]$, reducing the cost below the reference steady jet since a very similar convective heat transfer enhancement ($J_a$) is achieved at a fraction of the required mass flow rate for the action of the jets ($J_b$). Interestingly, the optimisation process features an initial convergence to a local minimum in just $5$ generations $\bm{\Xi}^\star_5 = [216\mathrm{Hz}, 216\mathrm{Hz}, 62.5\%, 62.5\%, 44^\circ]$, in which the optimised frequency and duty cycle of each valve are already converged as depicted in figure~\ref{fig:LGAC_progress}(d,e). After the $10^{th}$ generation, an additional improvement is achieved in the controller performance, which is directly related to a change in the phase, $\bm{\Xi}^\star_{10} = [216\mathrm{Hz}, 216\mathrm{Hz}, 62.5\%, 62.5\%, 30^\circ]$ (see figure~\ref{fig:LGAC_progress}f).
It must be remarked that the minimum value of the cost function $J$ is observed only in the last two generations. While this might open questions regarding possible further improvements with subsequent generations, it will be shown later in this section that the difference achieved in terms of $\mathrm{Nu}$ with the phase shift refinement of the last iterations is comparable to the measurement uncertainty of $\mathrm{Nu}$. For this reason, the training process is interrupted at the $10^{th}$ generation (as customary also in several other works reported in the literature) to achieve a good compromise between potential improvement, the uncertainty of the measurements, and testing time.

The heat transfer enhancement is the main driver of the learning process driven by LGAC as shown in figure~\ref{fig:LGAC_progress}(b). On the other hand, the value of $\tilde{J}_b$ is a direct measure of the mass flow required for a given individual and it can be approximated as the average value between $DC_1$ and $DC_2$. From the evolution of $\tilde{J}_b$ reported in figure~\ref{fig:LGAC_progress}(c), it is confirmed that the optimisation process converges to a very localised region of the solution space covering $60\%\leq DC \leq 70\%$ for both control laws. During the first $5$ generations, the algorithm progressively learns how to improve both $J_a$ and $J_b$, as shown from the unsorted distributions of individuals in figures~\ref{fig:LGAC_progress}(b,c) for $1\leq g \leq 5$; however, from $g=6$ on, the feasible solution space for the duty cycle is narrowed to a small interval, making the heat transfer augmentation the main driver of the optimisation process. 

\begin{figure}[t] 
    \centering\
    \includegraphics*[width=0.99\linewidth]{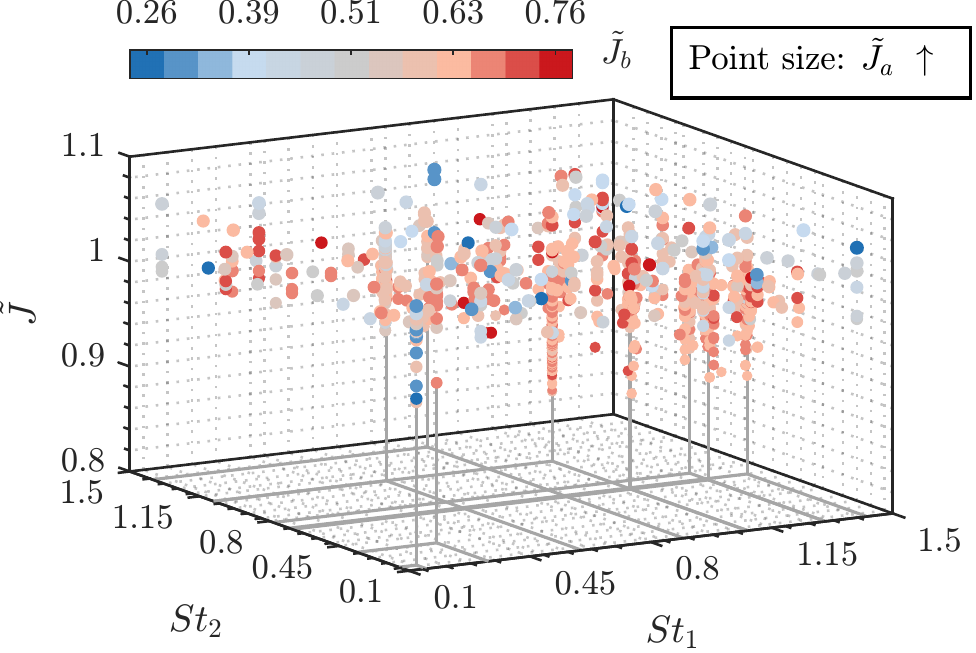}
    \caption{Distribution of controller performance as a function of actuation parameters. Each point represents an individual of the LGAC optimisation process. The colour indicates the penalty term $\tilde{J}_b$ and the size of the point grows with $\tilde{J}_a$.}
    \label{fig:param_interpretation}
\end{figure}

The previous discussion on the evolution of $J_a$ and $J_b$ during the optimisation process is supported by the parameter distribution depicted in figure~\ref{fig:param_interpretation}. Similarly to figure~\ref{fig:param_groups}, this representation of the data allows an understanding of the exploration and exploitation mechanisms of LGAC. The set of $925$ individuals is distributed over the whole feasible solution space. The figure highlights those clusters of points in which, for a fixed pair $[f_1,f_2]$, LGAC seeks for optima by varying the remaining control parameters. These phenomena appear in the form of the vertical distribution of individuals in which the total cost $J$ changes. It is worth noting that the minimisation of the cost $J$ is not dependent on the mass flow injection, namely $J_b$ when focusing on the vicinity of one of the preferred frequencies $f^*$. The cost is minimised for a fixed set of actuation frequencies and with almost unchanged duty cycles, meaning that the final control parameter to be optimised is the phase $\phi$. This is consistent with the parameter evolution reported in figures~\ref{fig:LGAC_progress}(d-f).

It is worth noting that it is a common practice to fix the phase among actuators to reduce the complexity of the problem of control optimisation. The feedback closed-loop control problems investigated by \citet{rabault2019JFM} and \citet{Castellanos2022LGPCvsRL} explore the utilisation of reinforcement learning and linear genetic programming control to reduce the drag of a 2D-cylinder at $\mathrm{Re}=100$ using two synthetic jets in phase opposition. Similarly, the work by \citet{minelli2020lgac}, based on linear genetic algorithms to find the optimised control of a bluff-body wake, adopts two in-phase jets. On the other hand, \citet{Wu2018jet} dodged the phase as an additional control parameter relying on the open-loop actuation literature, asserting that the phase becomes only relevant for a few frequency ratios, e.g. harmonic and sub-harmonic components of the mixing layer \citep{monkewitz1988phase}. On the opposite side, the recent study by \citet{Yu2021GA_drag_slot_jets} just focuses on the initial phase delay among six synthetic slot jets to reduce drag on a turbulent boundary later. Note, however, that the frequency and duty cycle of the periodic forcing are fixed so that each actuator is performing the same control action with a phase shift. Our results indicate that the optimal frequency and duty cycle are identified relatively early in the optimisation process, and that relevant performance improvement can be obtained by fine phase tuning.

\begin{figure*}[td] 
    \centering
    \includegraphics[width=0.95\linewidth]{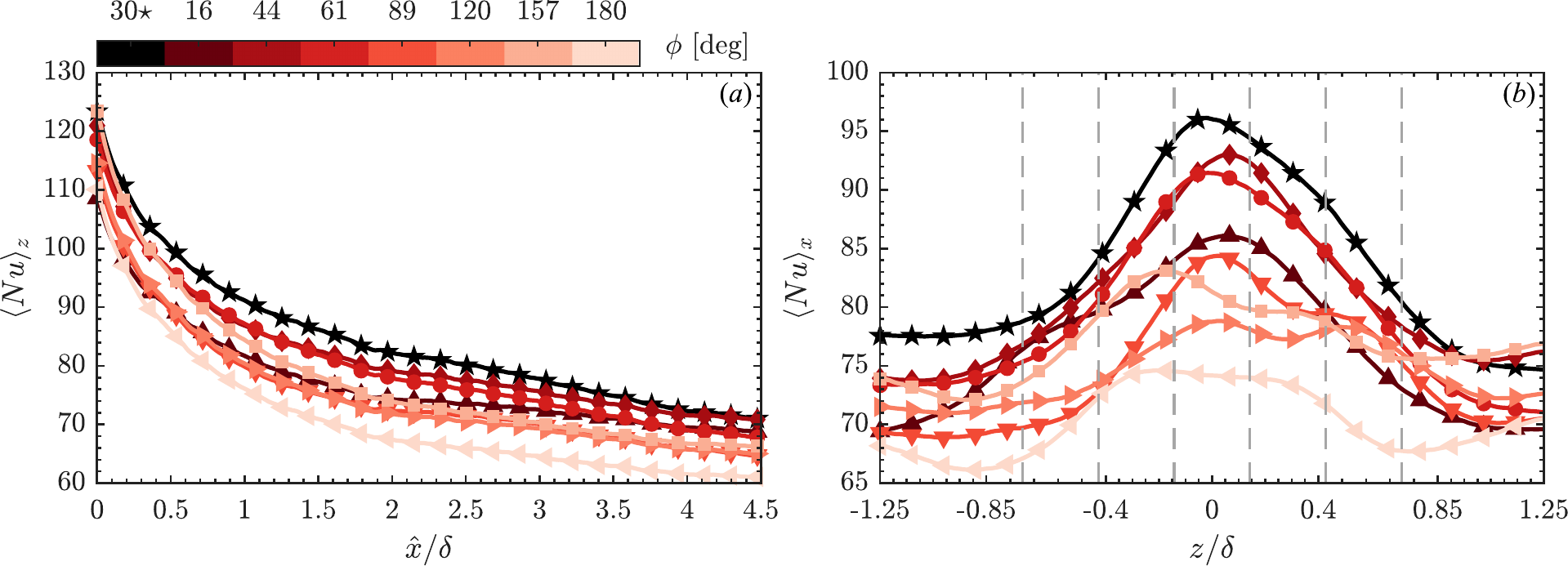}
    \caption{Nusselt profiles averaged along the spanwise (a) and streamwise (b) direction for the optimised combination of frequencies and duty cycles ($f_1 = f_2 =216\mathrm{Hz}$ and $DC_1 = DC_2 = 62.5\%$) for different phases. The best individual from the LGAC optimisation process is highlighted with the symbol $\star$.}
    \label{fig:Nuprofiles_phase}
\end{figure*}

Figure~\ref{fig:Nuprofiles_phase} depicts the streamwise and spanwise averaged profiles of Nusselt number ($\langle \mathrm{Nu} \rangle_x$ and $\langle \mathrm{Nu} \rangle_z$, respectively) for the different phases while maintaining the optimal combination of frequencies and duty cycles identified by LGAC. It must be remarked that this process is carried out outside of the population explored by LGAC, i.e. we are simply fixing all parameters of the control laws to those of the best individual except the phase. The phase of the best individual identified by LGAC, highlighted in black and with the symbol $\star$, exhibits the highest value of $\mathrm{Nu}$ both along the streamwise and spanwise directions over the analysed area downstream of the actuators. The performance of the actuators is considerably reduced when increasing $\phi$, until reaching the out-of-phase condition ($\phi = 180^\circ$) at which the performance is the worst. A particularly noteworthy effect of the phase on the $\mathrm{Nu}$ distribution is the induced spanwise asymmetry. The authors hypothesise that the change in phase between adjacent jets is promoting the translation of the induced structures along the spanwise direction as it will be later discussed in \S\ref{s:discussion}. When acting out of phase, the jets driven by $b_2$ cancel most of the effect caused by the adjacent jets driven by $b_1$. On the contrary, for the optimised phase $\phi^\star=30^\circ$, it seems that the jet-induced structures are convected towards the positive spanwise direction, hence yielding the asymmetric distribution of Nusselt number at the wall. This asymmetry changes sides depending on the phase, as shown in figure~\ref{fig:Nuprofiles_phase} for $\phi>90^\circ$.

\begin{figure*}[td] 
    \centering
    \includegraphics[width=0.95\linewidth]{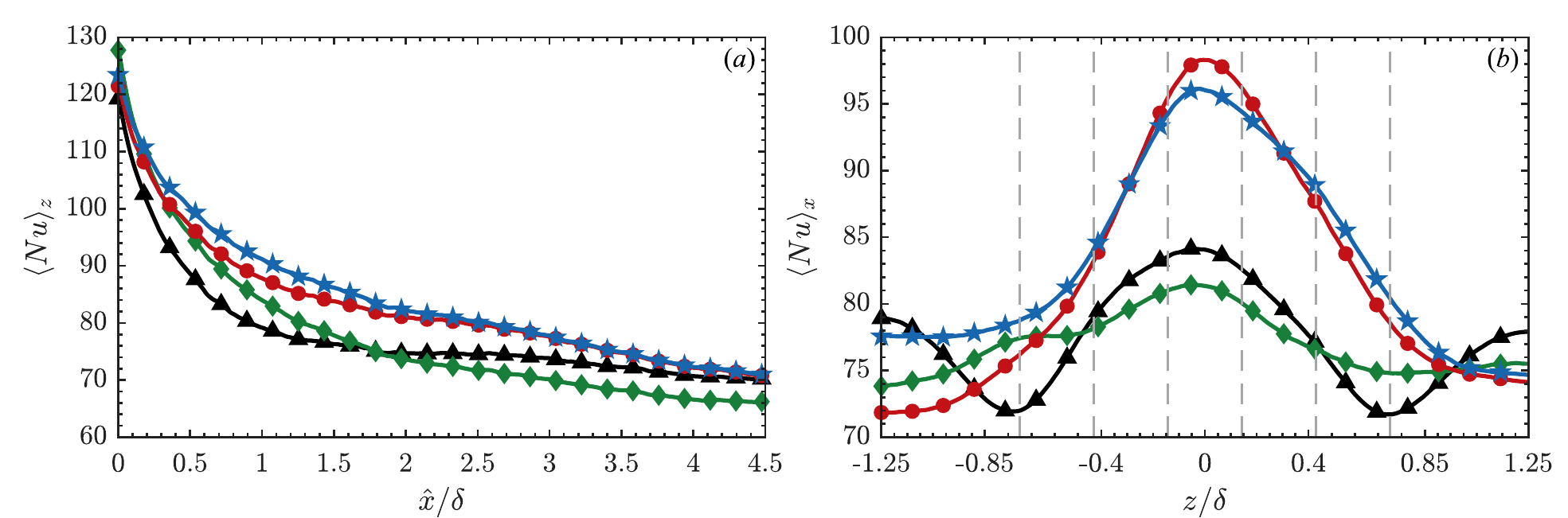}
    \caption{Nusselt profiles averaged along the streamwise (a) and spanwise (b) direction for the steady jet \sy{black}{t*}, $g = 1$ \sy{green_lgac}{d*}, $g = 5$ \sy{red_lgac}{o*} and $g = 10$ ({\color{blue_lgac}$\star$}).}
    \label{fig:Nuprofiles}
\end{figure*}

Finally, it is worth discussing the evolution of the Nusselt number during the optimisation process. Figure~\ref{fig:LGAC_progress}(g-i) shows the convergence of the Nusselt number distribution at the wall just downstream of the actuator, while figure~\ref{fig:Nuprofiles} illustrates the spanwise and streamwise averaged profiles of those respective fields. Even though the performance of the best individual after $g=1$ is better than the steady jet, the control law based on $\bm{\Xi}_1^\star$ produces an effect on the heat transfer distribution at the wall that is considerably less persistent along the streamwise direction and with a flatter profile along the spanwise component. For the following generations, the injection of mass is increased and a characteristic frequency of the system is found. The first conclusion is the tendency to asymmetric Nusselt distributions for the best strategies. The convective heat transfer increases towards the positive spanwise direction, suggesting a deviation of the flow that produces an improvement in the mixing capabilities within the TBL. The phase is the control parameter inducing the flow asymmetry, as previously observed in the evolution of the control parameters during the training process (figure~\ref{fig:LGAC_progress} and figure~\ref{fig:param_interpretation}) and the variation of Nusselt number profiles for different phases (figure~\ref{fig:Nuprofiles_phase}). This result, perhaps surprising at first observation, can be explained in terms of the asymmetry of the chosen configuration. The pairs are indeed formed using adjacent actuators, which results in asymmetric actuation with respect to the midline crossing the actuators in the streamwise direction. While this choice provides symmetric conditions with respect to the central pair of actuators, the overall actuation configuration becomes asymmetric.

\section{Discussion on the actuated flow}\label{s:discussion}
\begin{figure*}[t] 
    \centering\
    \includegraphics*[width=0.95\linewidth]{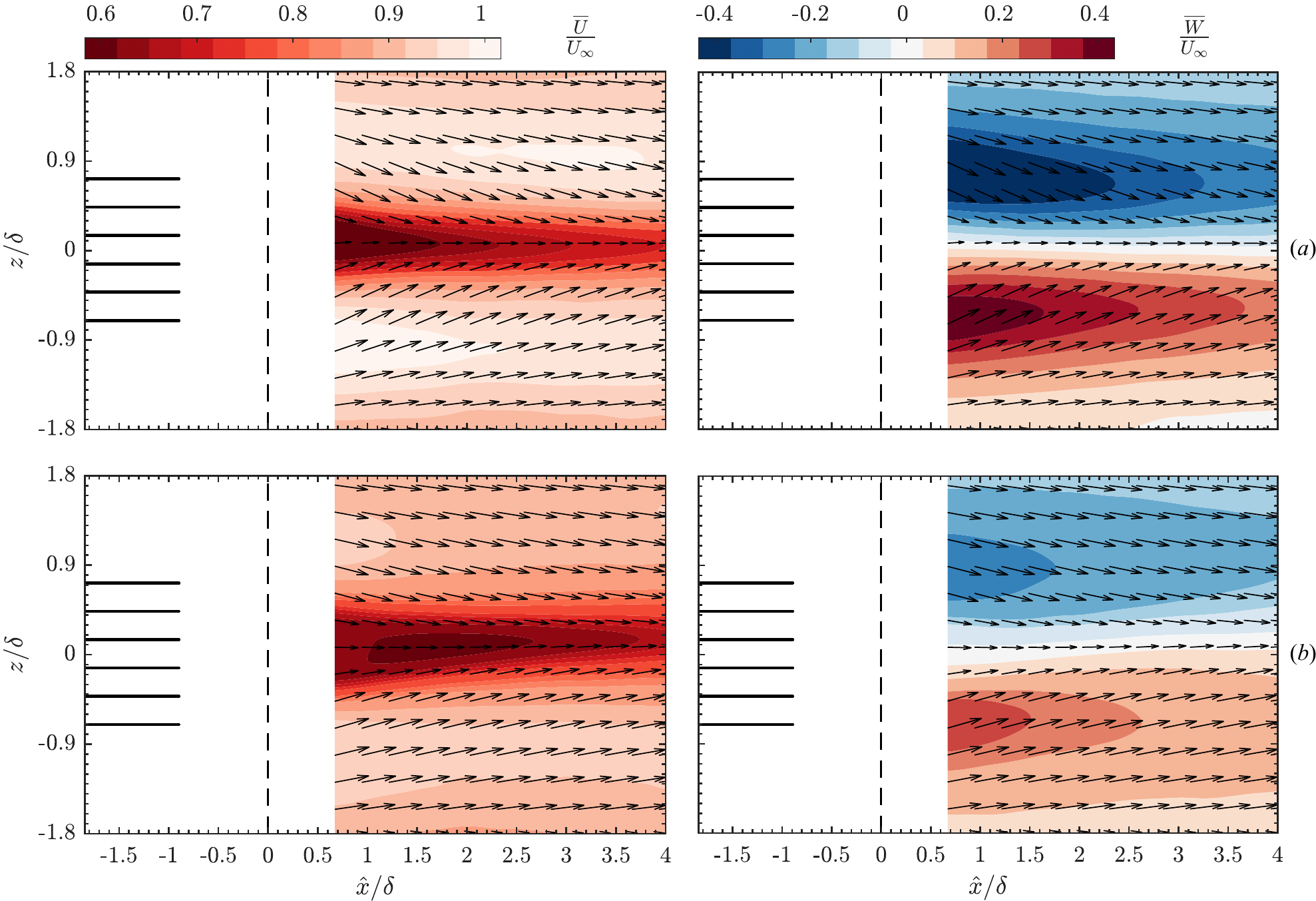}
    \caption{Average streamwise $\overline{U}$ and spanwise $\overline{W}$ velocity flow fields at $y^+ = 350$ for steady-jet actuation (a) and best individual from LGAC (b). The vector field shows the in-plane motion. Results are normalised with the freestream velocity $U_\infty$, and the location of the slots \lcap{-}{black} and $x_{s}$ \lcap{--}{black} are included for reference.}
    \label{fig:PIV}
\end{figure*}

The flow topology upon the action of the pulsed slot jets is discussed in this section. A comparison of the flow field induced by the optimised controller and the one from the steady-jet actuation is performed. PIV measurements are carried out in a wall-parallel plane at $y^+=350$. At this height, the mean streamwise velocity of the mean flow for the unperturbed TBL is $\approx 0.83U_\infty$.  This plane lies at the outer edge of the logarithmic layer, close to the transition to the wake of the TBL profile as shown in figure~\ref{fig:TBLprofile}. Figure~\ref{fig:PIV} depicts the velocity magnitude distribution with the superimposed vector field of the in-plane motion. The steady-jet flow is characterised by a symmetric flow field in which the flow tends to be entrained towards the mid-region. This determines an acceleration of the flow at the edges of the region equipped with slots. This seems consistent with the observed distribution of the average Nusselt number (figure \ref{fig:reference_Nu}), with a prolonged region of persistent heat transfer enhancement favoured by the entrained flow.

This flow organisation is in line with the observations of a recent study by \citet{puzu2019jet}, which investigates the heat transfer enhancement and the flow topology modifications of a round jet in crossflow with a similar velocity ratio as in the present study. They report that the jet stream divides the mainstream flow around the jet exit, merging back after a certain distance downstream. This merging favours upward mixing of the jet-induced and mainstream flows due to the appearance of counter-rotating vortex pairs at both sides of the jet. For the six-slot case, the mutual interaction of the induced flow by the internal jets occurs earlier than the merging of the main freestream flow; however, figure~\ref{fig:PIV} shows the entrainment of air towards the central region of the flow field leading to a substantial deceleration of the flow. The jets stream penetrates through the boundary layer lifting part of the mainstream upwards, while another part is deviated sideways forming a pair of horseshoe vortices at both sides of the actuators, which is responsible for the entrainment of fresh air from the main flow and, consequently, of the heat transfer enhancement. The fluid diverted sideways, namely $|z/\delta| > 0.7$, accelerates penetrating toward the central line. It is worth noting that the modification of the flow topology by the steady-jet actuation resembles that of a solid obstacle. In both cases, the mainstream diverts to the side of the actuator while detaching rolling vortices. These induce a shedding flow that is the predominant feature of the actuated flow as later described.

The optimised controller exhibits certain differential aspects in the topology compared to that of the steady-jet case. The in-plane motion and the spanwise velocity component in figure~\ref{fig:PIV}(b) confirm the entrainment of the mainstream from the sides of the actuators in a similar fashion to the steady jet, though less abrupt. The streamwise velocity, however, is rather different. The jet stream lifts up the mainstream reducing its speed in the central region just as in the steady-jet case but covering a wider area in the spanwise direction. Nonetheless, the velocity magnitude increases just on top of the jets while the sides remain almost unchanged with mild acceleration. This leads to hypothesising a different mechanism, in which the flow is not diverted by the blockage effect of the jet, but rather is energised by the pulsating jets. This explains the uniformity in the in-plane motion and the wider region of influence in the spanwise $\mathrm{Nu}$ profiles in figure~\ref{fig:Nuprofiles}. Moreover, it could also explain the higher persistence of the heat transfer enhancement with respect to the steady-jet case. In the latter, most of the energy is spent in the mutual interaction of the shedding flow and the reorganisation of the flow within the TBL while for the optimised controller, the energised flow ensures further downstream exchanging energy with the wall in form of thermal fluxes.

\begin{figure}[t] 
    \centering\
    \includegraphics*[width=0.99\linewidth]{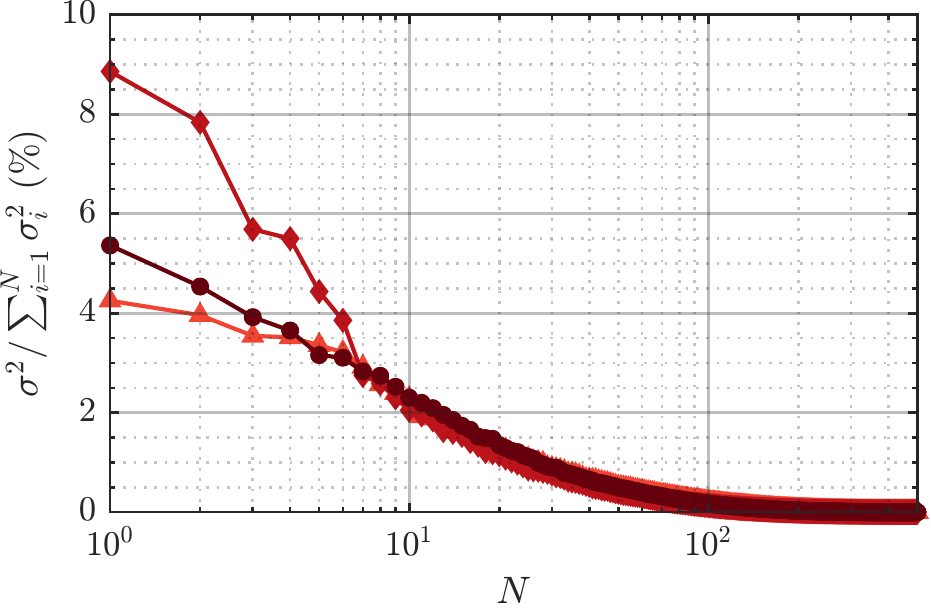}
    \caption{Spectrum of the POD modes for the unperturbed TBL \sy{red_pod_1}{t*}, steady-jet actuation \sy{red_pod_2}{d*}, and best individual from LGAC \sy{red_pod_3}{o*}}
    \label{fig:POD_spectrum}
\end{figure}

The eigenspectrum obtained from the POD analysis is shown in figure~\ref{fig:POD_spectrum}. The singular values represent the contribution of each mode in building the in-plane turbulent kinetic energy. The most relevant difference is observed in the first $6$ modes. Regarding the steady-jet actuation, the fraction of energy content in the first four modes is greater than for the other two cases, while the effect of the optimised controller concentrates in the first three modes, which are slightly more energetic than the unperturbed TBL. This similarity in terms of energy content suggests the efficiency of the actuation, which is directly related to a reduced injection of mass by the actuators. Conversely, the steady-jet actuation is considerably more energetic, which is the main reason behind the heat transfer enhancement in this case. These results conclude the capability of the optimised controller to outreach the performance of the steady jet, not only in terms of mass flow requirements but with a fraction of the energy contained within the altered flow.

\begin{figure*}[t] 
    \centering\
    \includegraphics*[width=0.99\linewidth]{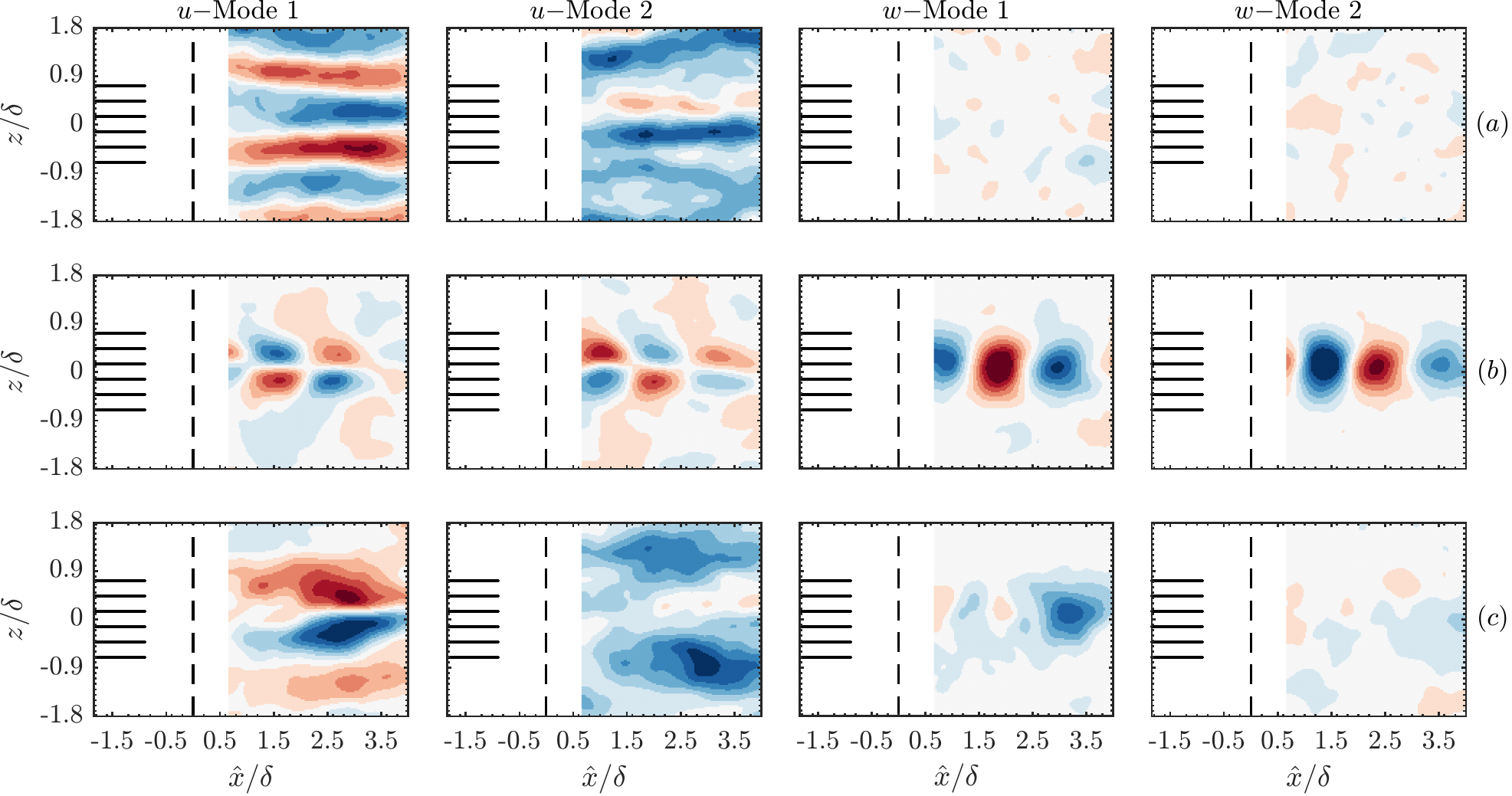}
    \caption{First two spatial modes of the streamwise $u$ and spanwise $w$ unsteady velocity component at $y^+ = 350$ for the unperturbed TBL (a), steady-jet actuation (b) and best individual from LGAC (c). The location of the slots \lcap{-}{black} and $x_{s}$ \lcap{--}{black} are included for reference.}
    \label{fig:POD}
\end{figure*}

The two first POD modes for the streamwise and spanwise unsteady velocity component ($u$ and $w$, respectively) are illustrated in figure~\ref{fig:POD} for the unperturbed TBL, the steady-jet actuation and the optimised controller. The $u-$modes of the unperturbed TBL describe a Fourier mode pair, given the homogeneity in the spanwise direction, while no relevant physical phenomena are reported in $w-$modes since most of the turbulent kinetic energy is related to the streamwise velocity fluctuations. For the steady-jet actuation, the POD modes resemble a classical shedding flow. The merged effect of the slot-jets in steady operation act in a similar fashion to an obstacle within the boundary layer, in agreement with \citet{Compton1992vorticesJICF}. This strong shedding motion confirms the high energy content of the first modes reported in the eigenspectrum from figure~\ref{fig:POD_spectrum}. Conversely, the first $u-$mode for the optimised controller illustrates a large-scale structure at both sides of the actuation area, while the lack of energy content complicates the interpretation of the remaining POD modes. These streaky large-scale structures resemble those of the unperturbed TBL, whereas covering a wider area and with greater intensity. The perturbed flow is somewhat shifted towards $z>0$, which coincides with a higher and asymmetric value of $\mathrm{Nu}$ for $z/\delta > 0.7$ (see figure~\ref{fig:Nuprofiles}). 

Aiming at interpreting the optimised set of control parameters $\Xi^*$, it is worth treating the frequency in its dimensionless form. The three characteristic frequencies found by LGAC during the optimisation process correspond to $\mathrm{St}^* \approx 0.70, 1.09~\text{and}~1.32$. From the evolution of the individuals over each generation, it becomes evident the relevance of $\mathrm{St}^\star = 1.09$ as the preferred frequency for the controller. Based on the definition of the Strouhal number in Eq.~\eqref{eq:St}, the best performance of the controller is achieved when the pulsation excites the near-wall turbulent structures. It could be argued that the best individual triggers the production of smaller, closer-to-the-wall eddies, in agreement with our previous findings suggesting that near-wall coherent eddies are a suitable candidate for enhancing heat transfer \citep{mallor2019modal}. This result is in agreement with the finding reported by \citet{Castellanos2022slotjet}, in which the preferred frequency to maximise heat transfer enhancement with a single spanwise-aligned slot jet corresponds to $\mathrm{St} \approx 1$. However, the appropriate scaling for configurations involving pulsed jets in crossflow is still an open research question \citep{Sau2010optJICF}, and the optimal pulsation frequency, if any, is apparatus dependant \citep{MCLOSKEY2002}.
The optimised value of the duty cycle is a compromise between high heat transfer enhancement and low power consumption. The phase, however, has shown a surprising relevance in the final stages of the optimisation process. The phase is responsible for flow symmetry since it delays the interaction of the adjacent jet streams. This delay may exacerbate or jeopardise the mechanisms in charge of heat transfer enhancement as it was discussed. The optimised value $\phi^\star = 30^\circ = \pi/6$ promotes a deviation of the induced structures in the spanwise direction, which affects both the flow topology and the convective heat transfer distribution at the wall. Further analysis based on time-resolved measurements is required to shed more light on this interesting phenomenon.

\section{Conclusions \label{s:Conclusions}}

The use of Linear Genetic Algorithm Control to maximise heat transfer enhancement in a turbulent boundary layer on a flat plate with pulsed slot jets is investigated experimentally. An optimisation problem is formulated, dividing the set of actuators into two groups with independent actuation: odd and even jets based on their spanwise position. The actuation features a periodic forcing that depends on the frequency, duty cycle and phase between the two groups of jets. After a learning process with 10 generations and 925 individuals, an optimised combination of control parameters is found,
\begin{align*}
    \begin{split}
        \bm{\Xi}^\star & = [\mathrm{St}_1,~\mathrm{St}_2,~DC_1,~DC_2,~\phi] \\
        & = [1.09,~1.09,~0.625,~0.625,~\pi/6]
    \end{split}
\end{align*}

The optimisation process is guided by convective heat transfer enhancement while maintaining low power consumption requirements. Linear Genetic Algorithm Control achieves a convective heat transfer enhancement of 17.5\% with respect to the unperturbed TBL, which is higher than the steady-jet actuation case, at a fraction of the power consumption. Since the control laws converge to the same frequency for both sets of actuators, the phase difference between them is a meaningful control parameter. The relevance of the phase difference between control laws is addressed, being $\phi$ the main driver of an asymmetry in the actuated flow field. The optimised phase $\phi^*$ shifts the jet-induced structures in the spanwise direction. The optimised value of the duty cycle is a compromise between high heat transfer enhancement and low power consumption.

Flow field measurements based on Particle Image Velocimetry and a Proper Orthogonal Decomposition analysis are used to interpret the flow topology upon actuation. For the steady-jet actuation, part of the flow is diverted upwards by the jet's stream and another part moves sideways. The flow at both sides of the actuators accelerates promoting the generation of shedding phenomena responsible for the entrainment towards the central region downstream of the actuator. For the optimised controller, the flow remains within the central area although with a considerable reduction in energy content. However, the optimised controller is more efficient in spreading the heat transfer uniformly in the spanwise direction and enduring for a long distance downstream. 

This work demonstrates that genetically-inspired algorithms are effective in optimising the active control of a turbulent boundary layer for heat transfer applications. The selection of the optimised controller is valid though within the extent of this work. The proposed study is subjected to several technical and fundamental assumptions associated with the formulation of the optimisation problem, the implementation of LGAC to an experimental plant or even the interpretations of results. Yet, this study demonstrates the capabilities of an advanced model-free optimisation algorithm to solve a rather complex flow control problem. The results encourage exploring the full potential of the machine-learning-based control for heat transfer applications by extending the search space, e.g. by including feedback from sensor signals or multi-frequency forcing. 

\newpage
\section*{Research data}
The research data from experiments are available in Zenodo repository \citep{Castellanos2023Zenodo}.

\section*{Acknowledgments}
The work has been supported by the project ARTURO, ref. PID2019-109717RB-I00/AEI/10.13039/501100011033, funded by the Spanish State Research Agency. \\
\noindent S. Discetti acknowledges the funding provided by the European Research Council (ERC) under the European Union’s Horizon 2020 research and innovation program (grant agreement No. 949085, NEXTFLOW). Views and opinions expressed are however those of the author(s) only and do not necessarily reflect those of the European Union or the European Research Council. Neither the European Union nor the granting authority can be held responsible for them. \\
\noindent Juan Alfaro, Miguel A. G\'{o}mez and Isaac Robledo are kindly acknowledged for their support during the preparation and development of the experimental campaign.
The authors also thank Guy Y. Cornejo Maceda for providing the first version of the LGAC software.

\bibliographystyle{model1-num-names}
\bibliography{bib}

\end{document}